\documentclass{aa}  %Use this if you want to change to two columns  

\usepackage{graphicx}
\usepackage{txfonts}
\usepackage{longtable}
\usepackage{subcaption}
\usepackage{float}
\usepackage{adjustbox}

\usepackage{placeins}           % useful with \FloatBarrier, to keep 
\usepackage{orcidlink}          % allow ORCID ids
\usepackage{hyperref}

\usepackage{geometry}
\usepackage{supertabular}                                             
\usepackage[table]{xcolor}
\usepackage{overpic}
\usepackage{dblfloatfix}

\usepackage{color}
\definecolor{red(new)}{RGB}{200, 50, 60}

\begin{document}

   \title{Orbital and stellar parameter determination of eclipsing $\beta$ Cep pulsators}

   %\subtitle{I. Overviewing the $\kappa$-mechanism}

\author{C. I. Eze \orcidlink{0000-0003-3119-0399} \inst{\ref{CAMK},\ref{UNN}} \fnmsep\thanks{Corresponding author} \and
        G. Handler \orcidlink{0000-0001-7756-1568} \inst{\ref{CAMK}} \and
        F. Kahraman Ali\c cavu\c s \orcidlink{0000-0002-9036-7476} \inst{\ref{Turkey1},\ref{Turkey2}} \and T. Pawar \orcidlink{0000-0002-0004-0569} \inst{\ref{Villanova}} \and P. De Cat \orcidlink{0000-0001-5419-2042} \inst{\ref{ROB}} }

\institute{Nicolaus Copernicus Astronomical Centre, Polish Academy of Sciences, Bartycka 18, PL-00-716 Warsaw, Poland \\
           \email{cheze@camk.edu.pl} \label{CAMK} \and
           Department of Physics and Astronomy, Faculty of Physical Sciences, University of Nigeria, Nsukka, Nigeria\label{UNN} \and
           \c Canakkale Onsekiz Mart University, Faculty of Sciences, Physics Department, TR-17100 \c Canakkale, T\"urkiye\label{Turkey1} \and
           \c Canakkale Onsekiz Mart University, Astrophysics Research Center and Ulupınar Observatory, TR-17100 \c Canakkale, T\"urkiye\label{Turkey2} \and Department of Astrophysics and Planetary Sciences, Villanova University, 800 East Lancaster Avenue, Villanova, PA 19085, USA \label{Villanova}  \and
           Royal Observatory of Belgium, Ringlaan 3, B-1180 Brussels, Belgium \label{ROB}
           }

  % \date{Received September 15, 1996; accepted March 16, 1997}

% \abstract{}{}{}{}{} 
% 5 {} token are mandatory
 
  \abstract
  % context heading (optional)
  {Eclipsing binaries are veritable laboratories for the study of stellar structure and evolution. The dynamical properties (e.g. mass and radius) inferred from eclipsing binary modelling coupled with asteroseismic masses offer a unique opportunity to probe stellar evolution models. Spectroscopy characterises the stellar envelope and plays a key role in estimating more accurate dynamical parameters, as it provides invaluable priors for the eclipsing light curve modelling.} 
  % aims heading (mandatory)
  {In this paper, thirteen eclipsing binaries with massive primary components are spectroscopically and photometrically analysed to derive the orbital and stellar parameters, and infer their possible evolutionary status.} 
  % methods heading (mandatory)
  {The systems were spectroscopically observed using SALT HRS, HERMES, CHIRON and a spectrograph at Skalnate Pleso Observatory (SPO), and photometrically observed using TESS. Radial velocities and atmospheric parameters are determined, binary systems are modelled, and their absolute parameters are derived using appropriate modelling tools. The rotational properties, circularisation, and synchronisation of the systems are also investigated, and the general properties of the sample are discussed.} 
   % results heading (mandatory)
  {The sample shows one slow rotator, two circularised and synchronised fast rotating systems, as well as five circular and five eccentric systems with asynchronous fast rotation. The systems are predominantly SB1 or SB2 with faint companions, resulting in dynamical mass estimates at $5$ -- $12\%$ precision and radius estimates at $1.5$ -- $3.9\%$ precision. The secondary components appear to cut across a wide mass spectrum ranging from low-mass to intermediate mass systems.  The eccentricity of the systems shows certain trends with the projected rotational velocity ($v \sin i$) although there is no universal correlation between them. Also, there could be circularisation without co-rotation or synchronisation even for circularised short-period close binary systems.}
  % {The conclusion ...}
  {}

\authorrunning{Eze et al. }
\titlerunning{Orbital and stellar parameter determination of eclipsing $\beta$ Cep pulsators }

\maketitle
%
%-------------------------------------------------------------------

\section{Introduction}\label{sec:introduction}

Stars are the building blocks of the universe. The study of their structure and evolution paves the way for a better understanding of our Galaxy and the universe in general. To investigate stars, their basic stellar parameters, such as mass, radius, and chemical composition, are probed. One of the observational methods for obtaining these parameters is spectroscopy, which accesses the envelope of the stars. Through the analysis of a star's spectra, one obtains its orbital elements and atmospheric solutions, and is able to derive the stellar parameters such as mass and radius. However, it has been observed that the masses of massive stars obtained from spectroscopy are different from the masses derived from stellar evolution models \citep{Herreroetal1992}. This results in the mass discrepancy problem, which is attributed to the underestimation of the convective core masses. The stellar cores are beyond the reach of spectroscopy. Asteroseismology, which is the study of stars via their oscillations, bridges this gap. It offers us the opportunity to probe the interiors of stars. 

Massive stars are predominantly found in binary and multiple systems \citep{Sanaetal2012, Sanaetal2014, Kobulnickyetal2014, SouthworthandBowman2022}, hence a substantial fraction is expected to also be found in eclipsing binaries.  By modelling eclipses, one obtains model-independent, precise stellar parameters (e.g. radius, mass), which serve as invaluable calibrators for stellar evolution theory \citep{Torresetal2010, Pedersenetal2019}. Dynamical masses derived from eclipsing binary modelling and asteroseismic masses offer unique opportunities to constrain the physics of stellar evolution models to treat the mass discrepancy problem between observation and models \citep{Tkachenkoetal2020}. To obtain more accurate stellar parameters from dynamical models, spectroscopy is needed, as it yields radial velocity semi-amplitudes, effective temperature, mass ratio (in case of double-lined spectroscopic binaries, SB2), and mass function (in case of single-lined spectroscopic binaries, SB1). These parameters serve as input priors for the eclipsing binary modelling and help to break the degeneracy among degenerate stellar parameters in the binary model.
   
Eclipsing binary stars serve as benchmarks for stellar parameters, especially masses and radii, and provide direct measurement of model-independent stellar parameters to unprecedented precision of about 1\% \citep{Southworth2012}. Unfortunately, only the stellar parameters of not more than 31 $\beta$ Cep stars in eclipsing binaries have been reported until date, with SB2 systems accounting for 14 out of the 31 objects \citep[see][]{Johnstonetal2019, LeeandHong2021, Buddingetal2021, Southworthetal2021, SouthworthandBowman2022, Miszudaetal2025, Cakirlietal2025, Cakirlietal2025B, Nazeetal2025}. \citet{Cakirlietal2025} and \citet{Cakirlietal2025B} account for over 57\% of the reported SB2s with \citet{SouthworthandBowman2022} and \citet{Nazeetal2025} reporting predominantly the properties of single-lined $\beta$ Cep eclipsing binaries. The orbital properties of V453 Cyg were first derived by \citet{Pavlovskietal2018} via the disentangling of the spectra of the components and were further used together with the TESS light curve in \citet{Southworthetal2020} to derive the stellar parameters.
   
Here, we seek to characterise, spectroscopically and photometrically a sample of $\beta$ Cep pulsators in eclipsing binaries to obtain their orbital and stellar parameters.  In Section 2, we describe the observation of the data. We also describe the spectroscopic analysis in Section 3 and, radial velocity and photometric light curve modelling in Section 4. We describe the absolute parameters in Section 5 and circularisation and synchronisation in Section 6. In Section 7, we discuss the results and write the conclusion in Section 8.

%---------------------------------------
\section{Observation}\label{sec:observation}

The systems analysed in this work were obtained from the catalogue of $\beta$ Cep pulsators in eclipsing binaries compiled by \citet{EzeandHandler2024b, EzeandHandler2024a}.  First, we preselected for immediate spectroscopic follow-up targets which are relatively bright (brighter than 12 mag in the V band) and having pulsation spectra with several significant pulsations that appear promising for asteroseismic analysis. The spectroscopic data of the sample were obtained using four different instruments.
These are the South African Large Telescope High Resolution \'{e}chelle Spectrograph (SALT HRS) \citep{Buckleyetal2006,Bramalletal2010,Bramalletal2012, Crawfordetal2010, Crauseetal2014} in South Africa, the HERMES spectrograph at the Mercator observatory \citep{Raskinetal2011} in La Palma, Spain, the CHIRON spectrograph in the SMARTS 1.5 m telescope in Chile operated by the SMARTS consortium \citep{Tokovininetal2013, Paredesetal2021}  and the spectrograph at the Skalnate Pleso observatory (SPO) \citep{BaudrandandBohm1992, Pribullaetal2024} in Slovakia. 

The ESO archive was also searched for possible sufficiently publicly available pipeline-reduced data of the targets in the catalogue. The results of the search were added to the pre-selected targets to obtain an intermediate sample. From the intermediate sample, we removed the systems in the ESO archive with spectroscopic analysis results in the literature for which we were unable to obtain sufficient spectra in our observation campaign with any of the other four instruments. We also removed from the intermediate sample, targets whose spectra we had obtained and whose secondary components are visibly clear in the spectra to be double-lined spectroscopic binaries (SB2). These targets are already being individually followed. Targets with insufficient data, bad or extremely poor data were also removed, resulting in a final sample of 13 targets and a total of 160 spectra obtained and analysed in this work. The data were obtained with the orbital cycle of each system in mind to be able to characterise their orbits. Brief observation logs of the targets are shown in Table \ref{tab:observation-details}.

\begin{figure*}[ht]
\centering
\includegraphics[width=0.75\linewidth]{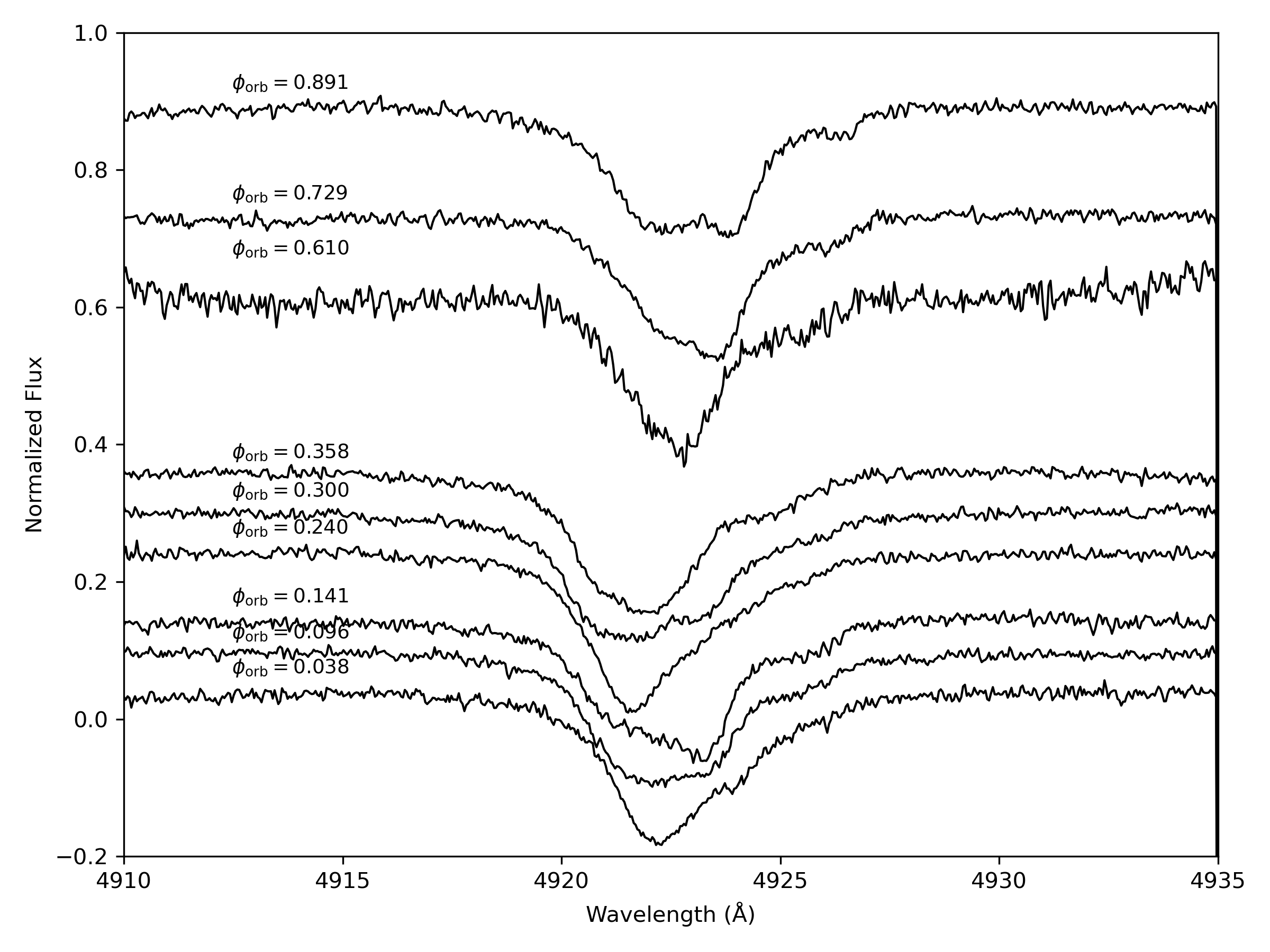}
\caption{A cut-out section of the normalised spectra of CPD$-$45 3109 showing the orbital radial velocity and pulsational line profile variations of He\,{\sc i}\,$\lambda4922$. The spectra are plotted according to the orbital phases.}\label{fig:line_profile_variability}
\end{figure*}

Using SALT HRS, we obtained, from $2020$ -- $2024$, 91 spectra for nine systems reported in this paper. 
SALT HRS is an efficient dual beam ($370$ -- $550$ nm and $550$ -- $890$ nm) fibre-fed, single-object \'{e}chelle spectrograph, which employs Volume Phase Holographic (VPH) gratings as cross disperser. The spectrograph delivers low resolution, R $\approx$ 14 000 (unsliced 500 $\mu$m diameter optical fibres), medium resolution, R $\approx$ 40 000 (sliced 500 $\mu$m fibres) and high resolution, R $\approx$ 65 000 (sliced 350 $\mu$m fibres).  
The different resolution modes of the SALT HRS were used with an exposure time that would give an optimal signal-to-noise ratio for the target at \ion{Si}{iii} 4567~$\AA$ at the stated resolution. The average signal-to-noise ratio (S/N) of our observation was about 100. Observing conditions from clear sky to thin cloud were allowed depending on the magnitude of the system and the cadences of the observations were decided on a case-by-case basis to reach the desired $S/N$. The SALT spectra were reduced using the SALT pipeline \citep{Kniazevetal2016, Kniazevetal2017}, and normalised using \textsc{IRAF/PYRAF} \citep{Tody1986}. An example of normalised spectra observed with SALT HRS is shown in Figure \ref{fig:line_profile_variability}. 

In 2022 and 2023, four systems reported here were also observed using the CHIRON spectrograph on the SMARTS 1.5-m telescope operated by the SMARTS consortium. CHIRON is a highly stable fibre-fed, cross-dispersed \'{e}chelle spectrograph whose multiple slit modes allow for spectral resolutions of about 26 000, 80 000, 95 000 and 130 000. It has a wavelength range of $440$ -- $880$ nm with 62 \'{e}chelle orders. 
34 spectra were obtained using the instrument at a resolution of about 26000. The observed spectra were partially reduced into 3D data cubes using the CHIRON pipeline. These partially reduced data were further reduced, barycentrically corrected, and normalised using \textsc{IRAF} tasks \citep{Tody1986} and \textsc{SUPPNET} \citep{Rozanskietal2022}, respectively.

The HERMES spectrograph was also used to obtain a total of 29 spectra for two systems in the sample. HERMES is a high-resolution spectrograph mounted on the 1.2-m Mercator telescope in La Palma, Canary Islands, Spain, with a resolution of 85 000. 
It observes within the wavelength range of $380$ -- $900$ nm. The observed spectra were reduced using the HERMES pipeline and subsequently normalised via spline fitting in \textsc{IRAF/PYRAF} continuum task.

In early 2023, we also obtained 6 spectra for two targets in the sample using the MUSICOS-type spectrograph at the 1.3-m (f/8.36) Nasmyth Cassegrain telescope at Skalnate Pleso Observatory (SPO) with a resolution of 35000. The spectrograph has a spectral range of $425$ -- $737.5$ nm in 56 \'{e}chelle orders. The spectra were reduced using \textsc{IRAF}, \textsc{FORTRAN} programmes and Linux shell scripts. 

The light curves used for the analysis are the Pre-search Data Conditioning Simple Aperture Photometry (PDCSAP) 2-min cadence light curves from the TESS mission \citep{Rickeretal2015} and downloaded from the MAST portal. The PDCSAP light curves  have been detrended of instrumental noise or long term trends using the PDC pipeline, while preserving intrinsic stellar variability. As a result, it is usually clearer and more suitable for our science goal. These higher cadence light curves sample the eclipse morphology more precisely than the longer cadence ones, allowing  better and more reliable fitting of light curves and stellar parameters.

%----------------------------------------------------------------------------------

\begin{table*}[ht]
\begin{center}
\caption{Observation details. NS is number of spectra, Instrument refers to telescope/spectrograph,  YO is year of observation, R is resolution and SWR is spectra wavelength range. The average signal-to-noise ratio (S/N) is 100. }\label{tab:observation-details}
\begin{tabular}[t]{cccccc}
\hline\hline
Star & NS  & Instrument& YO & R &  SWR (nm) \\
\hline\hline
CD$-38$ 4128 &11 & SALT HRS & $2019$ -- $2020$ &40000 & $370$ -- $550$ \\
%CD-51 9984 &10 & SALT HRS &2020 -2024 & 40000& 370-550  \\
CPD$-45$ 3109 &9 &SALT HRS & $2022$ -- $2023$ &40000 &  $370$ -- $550$ \\
HD 101838 &11 & SALT HRS & $2019$ -- $2020$ & 40000 &   $370$ -- $550$ \\
HD 108628 &10 & SALT HRS&2020 &40000 &      $370$ -- $550$ \\
HD 112026 &10 & CHIRON &2023 &26000  &   $440$ -- $880$ \\
HD 112485 & 13 &SALT HRS &2023 &40000 &  $370$ -- $550$ \\
HD 157400 & 7& SALT HRS&2022 &14000 &  $370$ -- $550$ \\
HD 254346 &19 &HERMES and SPO & $2018$ -- $2023$ &85000,35000 &  $380$ -- $900$, $425$ -- $737.5$  \\ 
HD 329379 &8 & SALT HRS& 2022&14000 &   $370$ -- $550$ \\
HD 339003 &17 & HERMES and SPO & $2018$ -- $2023$ &85000, 35000  &$380$ -- $900$, $425$ -- $737.5$ \\
HD 92741 &10 & CHIRON &2023 & 26000  &   $440$ -- $880$ \\
V1166 Cen &22 &SALT HRS and CHIRON &2023 &40000 & $370$ -- $550$, $440$ -- $880$  \\
V4386 Sgr &14 & SALT HRS and CHIRON& $2018$ -- $2023$ & 40000+, 26000 & $370$ -- $550$, $440$ -- $880$ \\
\hline\hline
\end{tabular}
\end{center}
\end{table*}

%--------------------------------------------------- One column table
  
%

%-----------------------------------------------------------------

\section{Spectral Analysis}

%-----------------------------------------------------------------

\begin{figure*}[ht]
\centering
\includegraphics[width=0.75\linewidth]{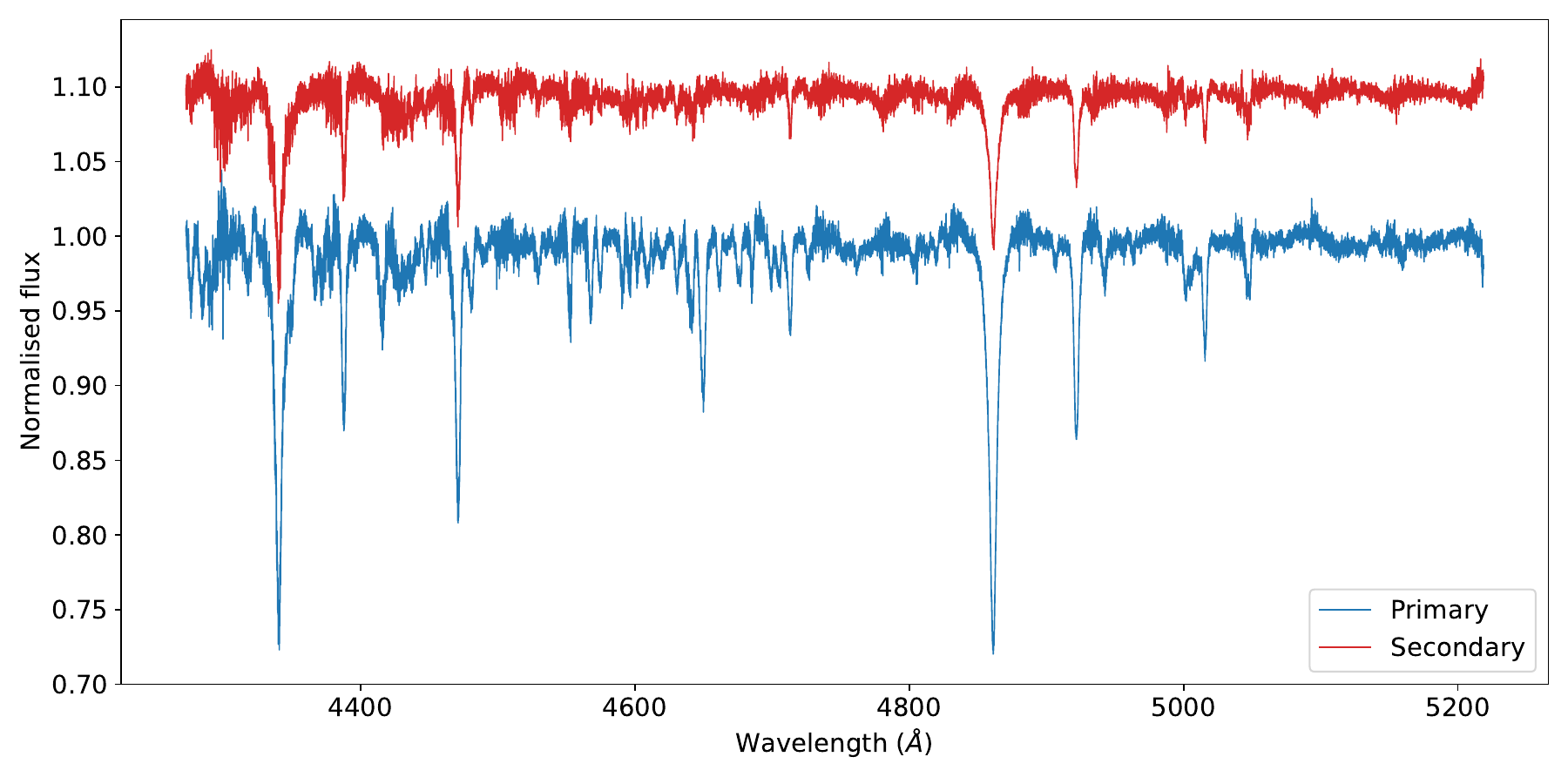}
\caption{An example figure showing the disentangled spectra of HD 329379.}\label{fig:disentangled_spectra}
\end{figure*}

%------------------------------------------------------------------

%\subsection{Line Profile variability}

%--------------------------------------------------------------------
%This ....

%------------------------------------------------------------------
\subsection{Orbital variability} \label{subsec:orbital-variability}

%---------------------------------------------------------------------
Binary motion induces radial velocity (RV) shifts in the spectra of binary systems. The RV is negative when the spectral lines of the orbiting star are blue-shifted towards us and positive when they recede from us. To understand the orbital dynamics of stars in binary systems, one has to first measure and model these RV signals to obtain the orbital elements, which are Keplerian orbital parameters that uniquely characterise the orbit of a star around the common centre of mass. Via the estimation of these parameters, which include eccentricity, radial velocity semi-amplitude, etc, the geometry, orientation, and velocity of the stars are inferred.
%In addition, modelling these orbital signals and removing them from the observed signals could pave the way for a possible investigation of the signatures of intrinsic variability such as pulsations \citep{Johnstonetal2021}. 
First, we measured RV values using the FXCOR task of \textsc{IRAF/PYRAF} \citep{Tody1986,2012ascl.soft07011S}. The IRAF FXCOR task uses a one-dimensional (1D) cross-correlation approach to match a synthetic spectrum, which serves as a template, with the observed spectrum to measure the RV shifts in the spectra. The relative RV of the observed spectrum is found as the correlation maximum, which is usually calculated as a function of the shift between the observed and template spectra in a one-dimensional cross-correlation algorithm \citep{TonryandDavis1979, MazehandZucker1994}. This method finds the RV even for extremely low S/N spectra owing to the fact that it uses all the information available in the spectra simultaneously \citep{MazehandZucker1994}. The synthetic spectra were generated in the \textsc{Spectrum} code \citep{GrayandCorbally1993} using appropriate ATLAS model atmospheres \citep{CastelliandKurucz2003} with effective temperature ($T_{\rm eff}$), surface gravity ($\log g$), micro-turbulence ($v_{\rm mic}$) and metallicity of $22000$ -- $30000\,\rm K$, $2.5$ -- $4.5$, $2.0$ -- $10.0 ~\rm km\,s^{-1}$ and 0.0, respectively, and hotiso isotope line list \citep{Gray1999}. Where necessary, the lines are broadened with appropriate $v \sin i$ values using the broadening function package with the \textsc{Spectrum} code. The preliminary choices of the appropriate $v \sin i$ for broadening, $T_{\rm eff}$ and $\log g$ for ATLAS model atmospheres, as well as  $v_{\rm mic}$ , are inferred from preliminary fits to the spectra using HANDY\footnote{https://github.com/RozanskiT/HANDY)}. Fixing the  $v_{\rm mic}$  to 2.0 $\rm km\,s^{-1}$ overestimates the $T_{\rm eff}$ of the stars, leading to an increase in the stellar mass \citep{Tkachenkoetal2020}. Hence, we used a rough preliminary estimate of  $v_{\rm mic}$  inferred from \textsc{HANDY} as one of the inputs for the synthesis of the template spectra. The RVs are measured from the spectra using a range of available unblended metal lines between $4200$ -- $7000~\AA$. Although using these ranges of lines gives more reliable and accurate RV values as they average over the RV of each independent line within the range, the errors are usually overestimated as they are the net errors from individual lines. This results in the large errors in the RV measurement reported in this paper, especially for fast rotators. \citet{Jilinskietal2006} also pointed out that the uncertainties in the RV measurements are significantly large for B-type stars with high $v \sin i$ values ($> 100\, \rm km\,s^{-1}$) due to the uncertainties in defining the line centre, whose position is also modified by pulsational line profile variations.  
The measured radial velocities are shown in Table \ref{appendix:measured-rvs}, with those of the SB2s derived from disentangling.

\begin{figure*}[ht]
\centering
\begin{subfigure}[t]{.33\textwidth}
\centering
\begin{overpic}[width=\linewidth]{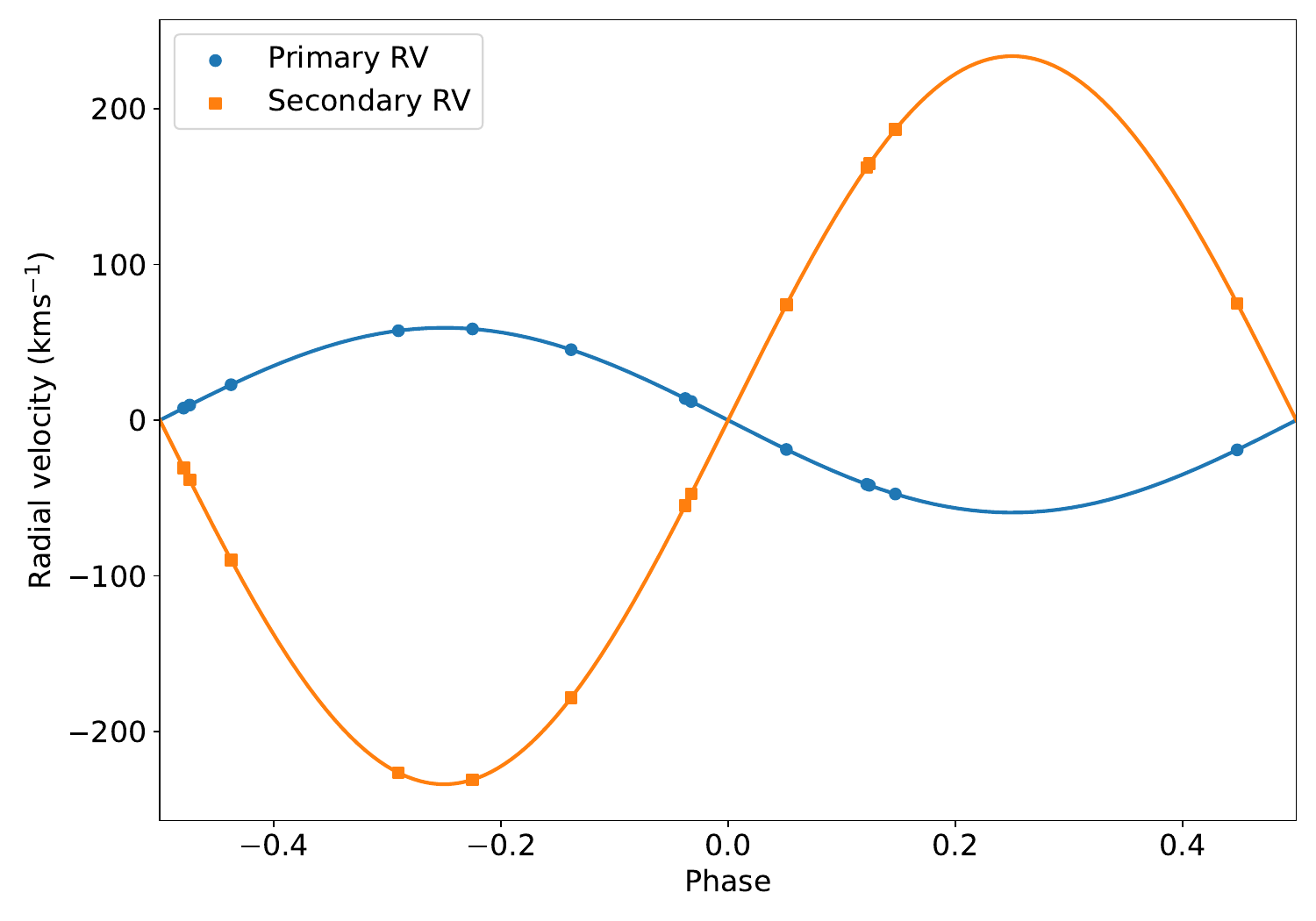}
  \put(70, 20){\small\bfseries HD 112485}
\end{overpic}
\caption{}\label{fig:hd1120}
\end{subfigure}
\begin{subfigure}[t]{.33\textwidth}
\centering
\begin{overpic}[width=\linewidth]{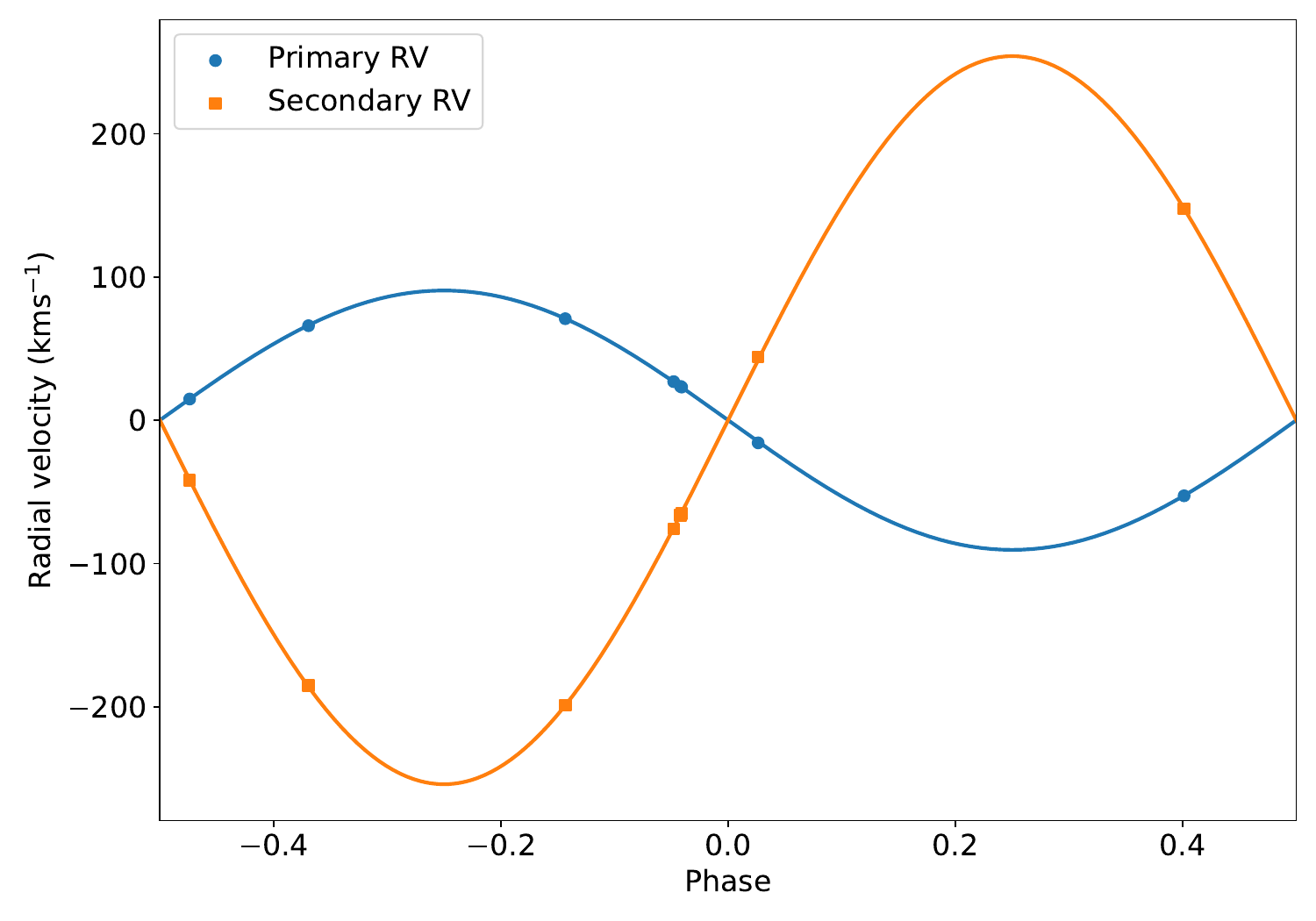}
  \put(70, 20){\small\bfseries HD 329379}
\end{overpic}
\caption{}\label{fig:hd1124}
\end{subfigure}
\begin{subfigure}[t]{.33\textwidth}
\centering
\begin{overpic}[width=\linewidth]{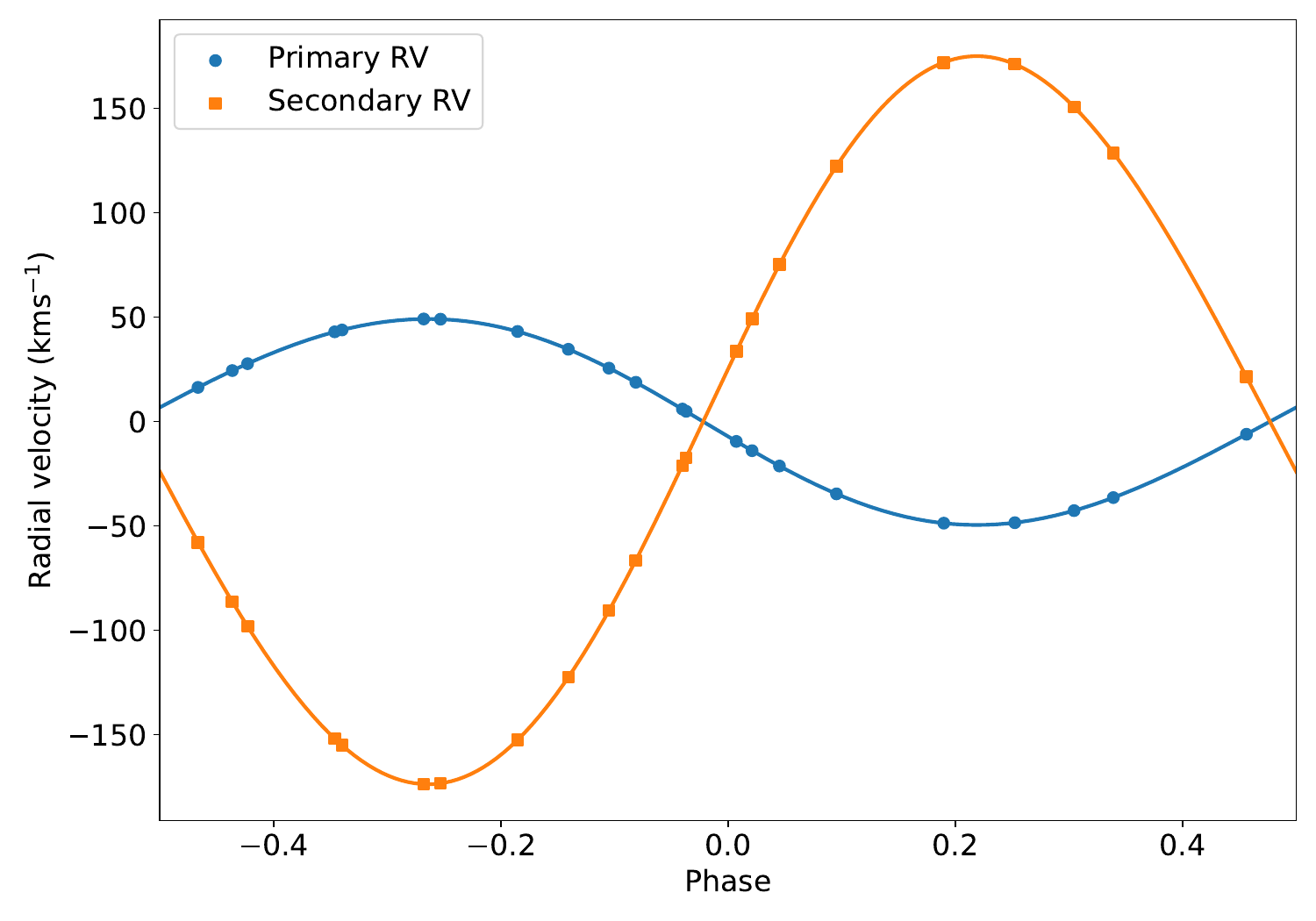}
  \put(20, 30){\small\bfseries V1166 Cen}
\end{overpic}
\caption{}\label{fig:hd108}
\end{subfigure}
%\begin{minipage}[t]{.4\textwidth}
\caption{Fitted radial velocity curves of SB2 systems in the sample obtained with disentangling. }\label{fig:SB2__orbital_fits}
%\end{minipage}
\end{figure*}

\subsection{Spectral disentangling}\label{sec:spectral_disentangling}
To search for faint companions to the targets in the sample, systems suspected  to be SB2s were subjected to spectral disentangling. These include HD 112026, HD 112485, HD 157400, HD 329379 and V1166 Cen. Using \textsc{FDBinary} (FD3) software \citep{Ilijicetal2004}, which is a C-based code designed for Fourier spectral disentangling of composite spectra, the orbital parameters and component spectra of these systems were determined over the wavelength range $4270$--$5220~\AA$. The radial velocities of the individual components were subsequently calculated from the optimised orbital solutions. The orbits of the secondary components of HD 112026 and HD 157400 were not reliably obtained due to their high eccentric nature and limited number of spectra, resulting in very large uncertainties  ($> 50~\rm km\,s^{-1}$) in $K_2$. Their $K_2$ values are suspected to be largely underestimated or overestimated for systems of their characteristics and spectral type.  Due to these limitations, their analyses are briefly described in Appendix \ref{appendix:sec_data_limitations} and a detailed follow up analysis is expected to be conducted on them when more spectra are obtained.

The RV points from disentangling are not independent measurements with known Gaussian errors such as those obtained for SB1. They are outputs of the same global spectral disentangling solution. As a result, their RV fits do not have error bars. Their scatter around the fitted RV curve can be tiny, because the disentangling procedure itself enforces an orbital solution very strongly. However, formal uncertainties for derived orbital solutions from disentangling are calculated at the spectral level and account for uncertainties in the entire disentangling process such as spectral noise, line blending, parameter correlations, global optimisation etc.

During the disentangling, the orbital period and the epoch derived from photometry are fixed. For circular orbits, the eccentricity and argument of periastron are fixed to 0 and 90, respectively. The disentangled spectra are scaled to the components' continuum fluxes using the normalised light contributions obtained for both components at first orbital quadrature (phase 0.25) from the preliminary modelled light curves. \textsc{Fdbinary} simultaneously optimises the orbital elements and the component spectra by minimising the residuals between the observed and reconstructed composite spectra. The parameter uncertainties from the disentangling are the formal $1\sigma$ errors reported by FDBinary, derived from the curvature of the objective function (i.e. the covariance matrix) at the best-fitting solution.  Figure \ref{fig:disentangled_spectra} shows the disentangled spectra of HD 329379 and Figure \ref{fig:SB2__orbital_fits} shows the RV fits or orbits of the SB2 systems derived from disentangling. The spectra of the components are subsequently used for the determination of the atmospheric parameters.   

%---------------------------------------------------------------------------
\subsection{Atmospheric Solutions}\label{subsec:atmospheric-solution}
%-----------------------------------------------------------------------
To derive the atmospheric parameters of the stars, we used barycentrically corrected and normalised spectra together with synthetic model atmospheres. Depending on the spectral type of $\beta$ Cephei stars \citep{EzeandHandler2024b}, we employed the non-local thermodynamic equilibrium (NLTE) Tlusty BSTAR2006 grids \citep{LanzandHubeny2007,HubenyandLanz2017} which are more sensitive to determine the atmospheric parameters of stars with high $T_{\rm eff}$ values \citep[$\gtrsim$\,15000~\rm K][]{2007ApJS..169...83L}. The NLTE grids cover the parameter ranges $15000$\,--\,$30000$\,K in $T_{\rm eff}$ in 1000\,K step and $1.75$\,--\,$4.75$\,dex in $\log g$ with 0.25\,dex step. Whenever necessary, the atmospheric parameters of synthetic spectra were linearly interpolated between grid points.

When multiple spectra were available for a given star (in case of SB1 systems), an average spectrum was constructed, whereas the disentangled spectra of the two components were analysed separately in the case of SB2. For stars exhibiting line-profile variations, this averaging procedure typically improved the stability of the atmospheric parameter determination. Before averaging, one spectrum was adopted as a reference, and all others were shifted to match its radial-velocity position to ensure proper wavelength alignment. By this method, we also increase the S/N ratio for the  examination target. 

During the analysis, first, the projected rotational velocities ($v\sin i$) were measured using the profile-fitting technique of \citet{Gray2005} before the determination of atmospheric parameters. Limb-darkened and instrumentally broadened synthetic spectra were compared to the observed He\,{\sc i} lines (e.g. $\lambda4387$, $\lambda4922$) in steps of $1~{\rm km\,s^{-1}}$. The uncertainties of the derived $v\sin i$ were derived considering the 1$\sigma$ level.

\begin{table}[ht]
\begin{center}
\caption{Atmospheric solutions of the stars.}\label{tab:atmospheric-solutions}
\setlength{\tabcolsep}{2.5pt}   % default is about 6 pt
\begin{tabular}[t]{lccr}
\hline\hline
Star & $T_{\rm eff}$ (K)  & $\log g$ & $v \sin i\, (km\,s^{-1})$   \\
\hline\hline
CD$-38$ 4128	&	21250(700)	&	3.7(1)	&	15(2)	\\
CPD$-45$ 3109	&	21800(800)	&	3.57(10)	&	134(5)	\\
HD101838	&	27000(1400)	&	3.60(14)	&	175(6)	\\
HD 108628	&	24500(1500)	&	3.93(11)	&	157(7)	\\
HD112026	&	28000(1500)	&	3.5(2)	&	60(4)	\\
HD 112485$_{1}$	&	25500(1500)	&	3.9(2)	&	176(5)	\\
HD 112485$_{2}$	&	-	&	-&		-\\
HD 157400	&	23500(1100)	&	3.95(13)	&	203(9)	\\
HD 254346	&	25000(1250)	&	3.83(14)	&	180(16)	\\
HD 329379$_{1}$	&	25000(2000)	&	3.75(20)	&	150(4)\\
HD 329379$_{2}$	&	15000(2000)	&3.7(3)	&	160(10)\\
HD 339003	&	28700(1100)	&	3.82(12)	&	214(12)	\\
HD 92741	&		-&		-&	95(5)	\\
V1166Cen$_{1}$	&	26000(2000)	&	3.9(2)	&	251(7)\\
V1166Cen$_{2}$	&	-&-	&	-\\
V4386 Sgr	&	23800(700)	&	3.63(13)	&	85(3)	\\
\hline\hline
\end{tabular}
\end{center}
\end{table}
%-------------------------------------------------------------------------

\begin{figure*}
    \centering
   \includegraphics[height=8cm, width=0.60\linewidth]{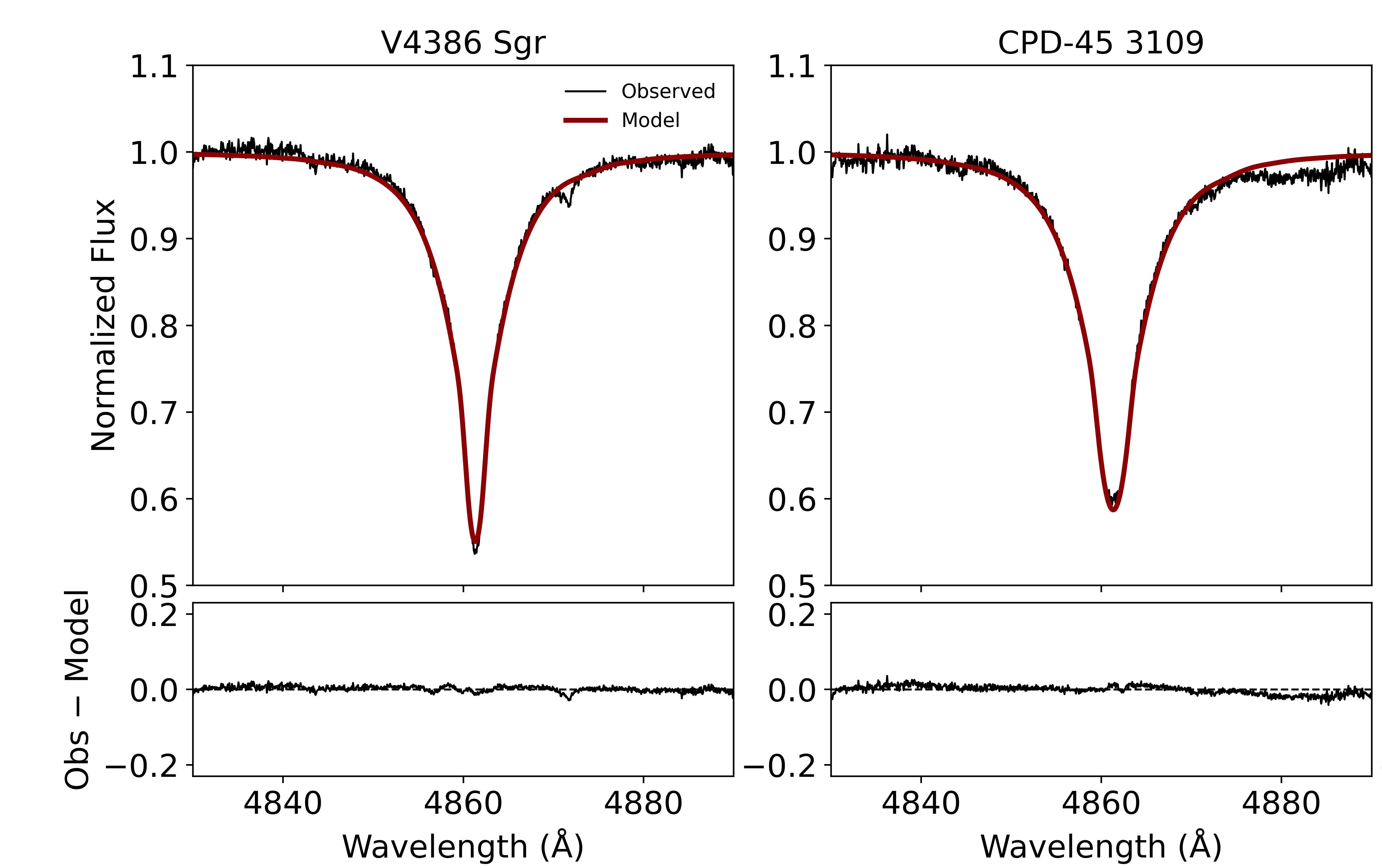}
   \includegraphics[height=7.8cm, width=0.28\linewidth]{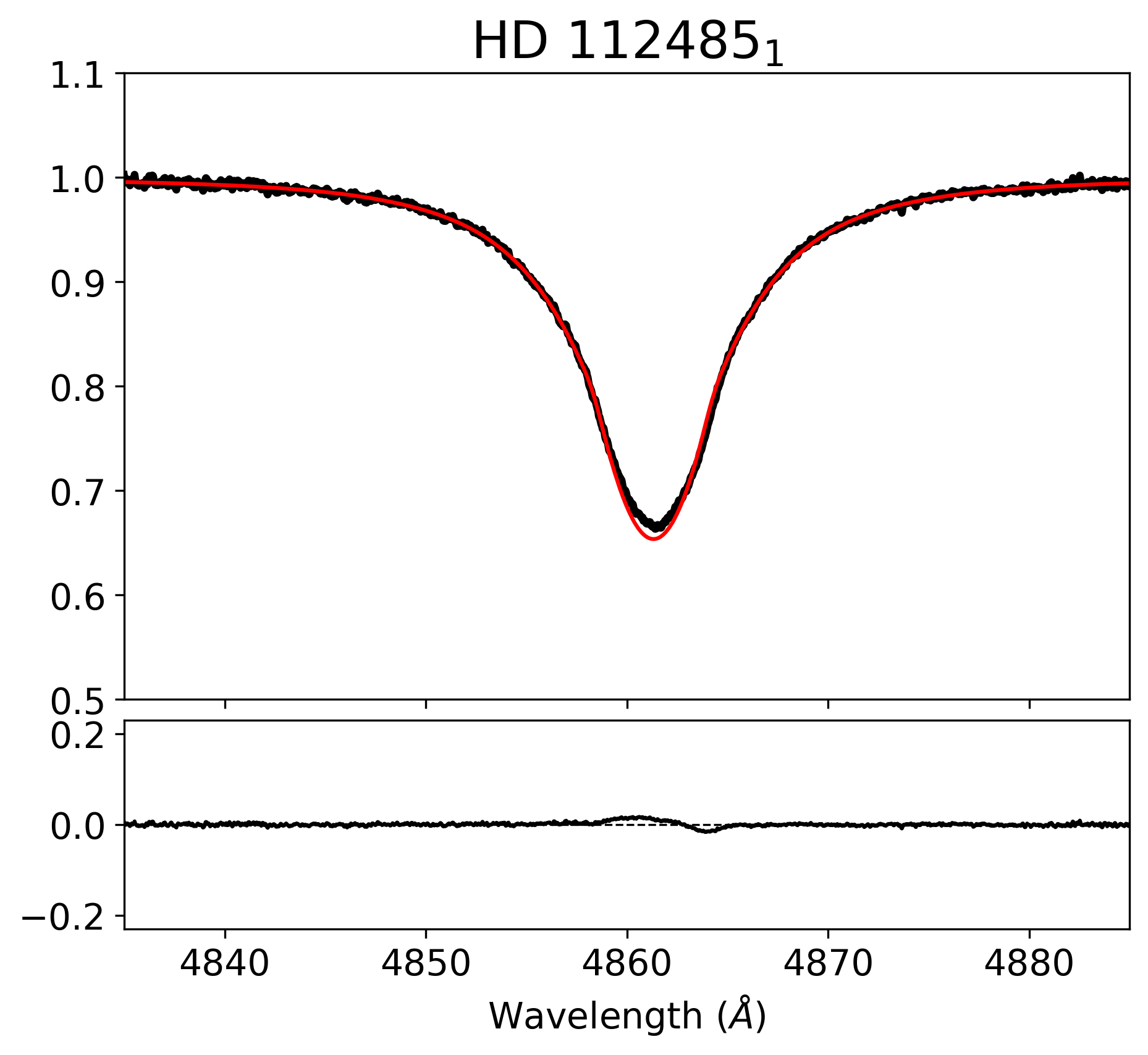}
    \caption{Upper panels: The consistency between the observed and the derived atmospheric models. Lower panels: residuals.}
    \label{fig:obs_model_consistency}
\end{figure*}

The $T_{\rm eff}$ and $\log g$ were determined by fitting the observed hydrogen Balmer and He\,{\sc i}  line profiles with synthetic ones, minimising the difference between the observed and synthetic lines following the procedure described by \citet{Catanzaroetal2004}. The uncertainties in $T_{\rm eff}$ and $\log g$  from each of the lines correspond to the 1$\sigma$ tolerance in the goodness-of-fit metric.  The values of $T_{\rm eff}$ and $\log g$  reported for the systems in Table \ref{tab:atmospheric-solutions} are the mean values of the measurements from individual  hydrogen Balmer and He\,{\sc i} lines, excluding lines that appear to be too strong or too weak, and their uncertainties are derived from standard error propagation on the mean. The measurements for each of the lines are shown in Table \ref{tab:Teffandlogg-from-lines}. For the stars, HD 92741 and HD 112026, the hydrogen lines in the CHIRON spectra were affected by artefacts; therefore, the atmospheric parameters were derived using helium lines. However, due to the presence of artefacts and significant line-profile variations, the uncertainties of the derived atmospheric parameters are relatively large, such that we had to discard the measurements for HD 92741. The spectra of the secondary components of HD~112485 and V1166 Cen have low signal-to-noise ratios and are too diluted by the low total secondary light contribution to yield reliable atmospheric parameters. 
The consistency between the observed and the derived atmospheric models are shown, for example in Fig. \ref{fig:obs_model_consistency} for three systems.

\section{Binary Modelling}\label{sec:binary_modelling}
The stellar parameters of the targets are determined using the combined strengths of the radial velocities and high precision TESS photometric light curve \citep{Rickeretal2015}. The PDCSAP light curves are used for the analysis. Where long term trends still exist in the data, we detrended the light curves by polynomial fitting and removed the long term trend from the data before modelling. We adopted different modelling routines for the the double- and single-lined spectroscopic binaries in our sample. 

\subsection{Double-lined spectroscopic binaries (SB2s)}\label{sec:SB2}
In the case of SB2, a joint modelling of the light and radial velocity curves was performed using \textsc{Wilson-Devinney} (WD) code \citep{WilsonandDevinney1971, Wilson1979, Wilson1990, Wilson2008, Wilsonetal2020} under a python framework \textsc{PYWD2015} \citep{GuzelandOzdarcan2020}. \textsc{PYWD2015} provides a graphical user interface and makes working with the WD code more user friendly. 

The radial velocities and the normalised detrended light curves of the targets are first phase-folded using the orbital period and epoch derived from the light curve using \textsc{JKTEBOP} \citep{Southworthetal2004}. Using mode 2 of the WD code, which was developed for detached eclipsing binaries, with the phase-folded light and radial velocity curves, a joint modelling was performed. During the modelling, the orbital period and the epoch are fixed to their respective values. The albedos and gravity darkening are fixed to 1.0, which is consistent with stars having radiative envelopes \cite[e.g.][]{vonZeipel1924}. The stellar atmosphere treatment available in the Wilson–Devinney code (IFAT = 1) was adopted, and logarithmic limb-darkening coefficients were automatically interpolated from the tables of \citet{vanHamme1993}. The grid size is set to N1 = N2 = 30 and NOISE = 1. The primary $T_{\rm eff}$ is also fixed to its value. 

\begin{figure*}[ht]
\centering
\begin{subfigure}[t]{.38\textwidth}
\centering
\begin{overpic}[width=\linewidth]{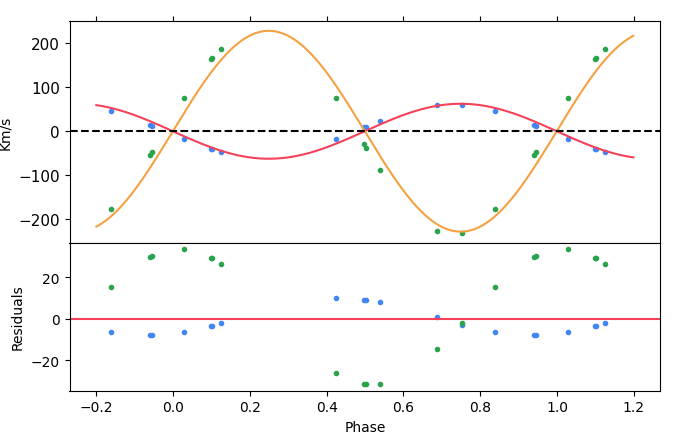}
  \put(52, 52){\small\bfseries HD 112485}
\end{overpic}
\caption{}\label{fig:cpd45}
\end{subfigure}
\begin{subfigure}[t]{.61\textwidth}
\centering
\begin{overpic}[width=\linewidth]{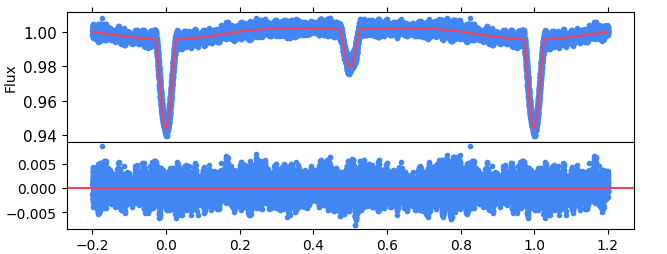}
  \put(40, 20){\small\bfseries HD 112485}
\end{overpic}
\caption{}\label{fig:hd1120}
\end{subfigure}

\medskip

\begin{subfigure}[t]{.38\textwidth}
\centering
\begin{overpic}[width=\linewidth]{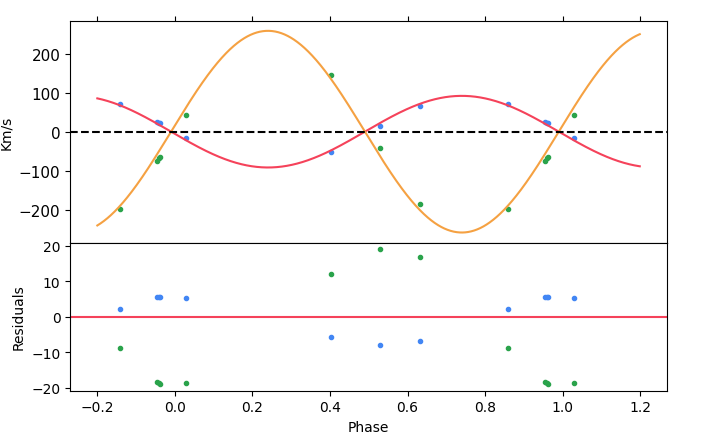}
  \put(53, 52){\small\bfseries HD 329379}
\end{overpic}
\caption{}\label{fig:hd101}
\end{subfigure}
\begin{subfigure}[t]{.60\textwidth}
\centering
\begin{overpic}[width=\linewidth]{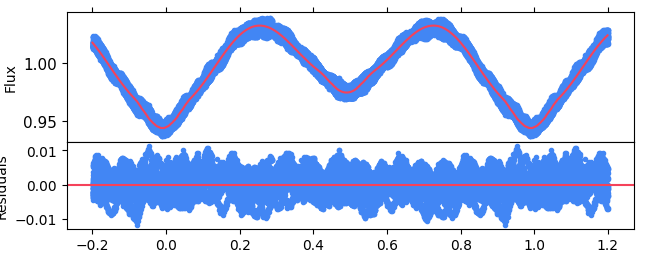}
  \put(40, 18){\small\bfseries HD 329379}
\end{overpic}
\caption{}\label{fig:hd1124}
\end{subfigure}

\medskip

\begin{subfigure}[t]{.38\textwidth}
\centering
\begin{overpic}[width=\linewidth]{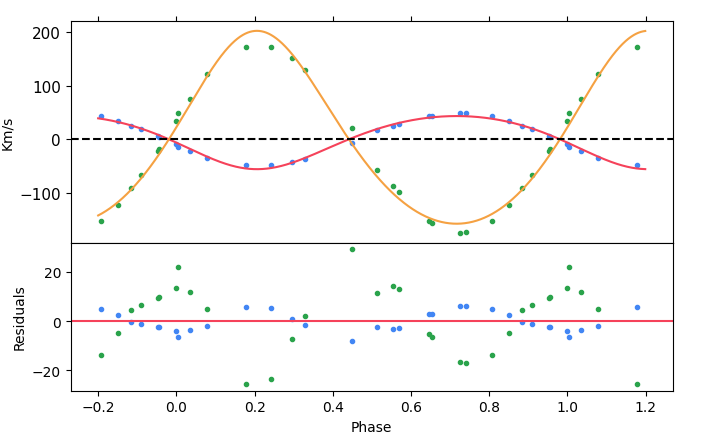}
  \put(30, 30){\small\bfseries V1166 Cen}
\end{overpic}
\caption{}\label{fig:hd108}
\end{subfigure}
\begin{subfigure}[t]{.60\textwidth}
\centering
\begin{overpic}[width=\linewidth]{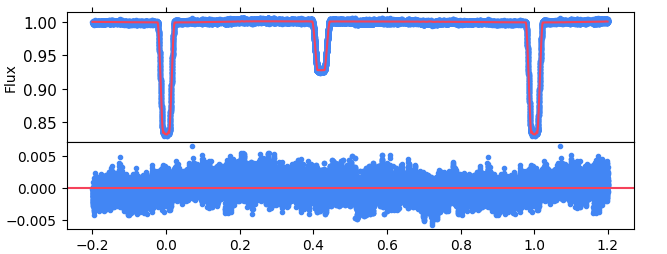}
  \put(40, 20){\small\bfseries V1166 Cen}
\end{overpic}
\caption{}\label{fig:hd157}
\end{subfigure}
%\begin{minipage}[t]{.4\textwidth}
\caption{Fitted radial velocity  and light curves of SB2 stars in the sample. }\label{fig:SB2_fits}
%\end{minipage}
\end{figure*}

We first manually initialised the model in the LC routine using the spectroscopic values until an approximate model is obtained. The physical parameters of the systems are adjusted in the differential correction (DC) routine of the WD code in a controlled sequential manner to ensure numerical stability. The parameters are adjusted until the adjustment is less than the standard errors. While keeping the semi-major axis ($a$) and mass ratio ($q$) to their spectroscopic values, the phase shift, the systemic velocity $\nu_{\gamma}$, the eccentricity $e$, and argument of periastron $\omega$ are first adjusted, then followed by the adjustment of the geometry parameters: inclination $i$, surface potentials $\Omega_{1}$ and $\Omega_{2}$. The temperature and the luminosity parameters: $T_{2}$  and $L_{1}$ are then adjusted followed by the adjustment of the third light $l_{3}$. Finally, the model is refined by the adjustment of the phase shift, $\nu_{\gamma}$,  $e$, $\omega$, $a$, $q$, $i$, $\Omega_{1}$, $\Omega_{2}$, $T_{2}$ and $L_{1}$, and sometimes  $l_{3}$. The fitted  light and radial velocity curves of the SB2 systems are shown in Figure \ref{fig:SB2_fits} and the physical parameters are shown in Table \ref{tab:SB2_parameters}. Note that in systems with circular orbits, $e$ and  $\omega$ remain fixed at their values of 0 and 90$^{0}$, respectively. 

The uncertainties in the derived absolute parameters are estimated using our own python implementation of  the Monte Carlo propagation of the covariance matrix of the final Wilson–Devinney solution. The adjusted parameters, their formal standard errors, and correlation coefficients were extracted from the final differential-corrections output (dcout.active). The covariance matrix was reconstructed as $C_{ij} = \rho_{ij}\,\sigma_i\,\sigma_j$, where $C_{ij}$, $\rho_{ij}$, and $\sigma_i$ denote the covariance, correlation coefficient, and standard deviation of the fitted parameters, respectively. Parameter vectors were generated by sampling from a multivariate normal distribution defined by the final WD parameter estimates and their covariance matrix, and the fixed parameters during the WD solution were assigned either external uncertainties or zero uncertainty, as appropriate, and were not included among the sampled WD parameters.

The volume-equivalent fractional radii were obtained from a local WD grid constructed only in the adjusted parameters controlling the stellar geometry, with all fixed parameters held constant. The WD lc program was evaluated only at the grid points, and the resulting radii were interpolated for each Monte Carlo realization. Absolute masses, radii, surface gravities, luminosities, and bolometric magnitudes were then calculated for every realization. Final parameter values and uncertainties were taken as the posterior medians and the 16th and 84th percentiles, respectively. This procedure propagates the covariance of the final WD solution into the derived stellar parameters through Monte Carlo sampling, but does not constitute a full forward-model MCMC exploration of parameter space.

\begin{table*}[ht]
\begin{center}
\caption{Orbital and physical parameters of the newly identified double-lined spectroscopic binary (SB2) systems fitted with PYWD2015."f" denotes fixed parameters. }
\label{tab:SB2_parameters}
\centering
\renewcommand{\arraystretch}{1.4}  % Default is 1.0
\begin{tabular}[t]{cccc}
\hline\hline
Parameters & HD 112485 & HD 329379 & V1166 Cen \\
\hline
$P$ (d)$^{f}$                         & 5.3720 & 2.2465 &13.4197  \\
$T_{0}$ (BJD$-2457000$)$^{f}$        & 1605.06500 & 3100.10000 & 1612.15384 \\
$e$                                  & 0 & 0 & 0.12586(4) \\
$\omega$ ($^{\circ}$)                & 90 & 90 &172.654(2)  \\
$K_{1,2}$ (km\,s$^{-1}$)             &62(3)/228(9)  & 92(4)/260(10) &  49.5(9)/179.5(3.4)\\
$q~(M_{2}/M_{1})$                    & 0.274(4) & 0.3551(2)& 0.2754(4) \\
$a$ ($R_{\odot}$)                & 31.44(1.26)  & 20.7(8) &  60.25(1.15)\\
$i$ ($^{\circ}$)                     & 78.60(5)	 & 49.13(8) & 89.60(2) \\
$T_{\rm eff,1,2}$ (K)                & 25500$^{f}$/13190(530) & 25000$^{f}$/16805(673) &26000$^{f}$/15612(601) \\
$\Omega_{1,2}$                       &  5.59(2)/5.07(4)& 2.772(4)/2.5874(9)	 & 9.952(5)/7.93(1) \\
$A_{1,2}$$^{f}$                            & 1.0 & 1.0 & 1.0 \\
$g_{1,2}$$^{f}$                            & 1.0 & 1.0 &  1.0\\
$X_{1,2}^{c}$                        & 0.331 /0.381 &0.347/0.355  & 0.330/0.346 \\
$y_{1,2}^{d}$                        & 0.207/0.203 & 0.219 /0.198 & 0.208/0.188 \\
$l_{1}/(l_{1}+l_{2})$                & 0.947(2) & 0.8016(9) & 0.92848(5) \\
$l_{3}$ (\%)                              & - & - & 0.0008(10) \\
$r_{1,2~pole}$ & 0.1878(7)/0.074(1)&   0.4084(7)/0.2733(3) & 0.10378(5)/0.04290(8) \\
$r_{1,2~point}$ &0.1892(7)/0.074(1) & 0.471(1)/0.383(6)  & 0.10394(5)/0.04294(8)  \\
$r_{1,2~side}$ &0.1886(7)/0.074(1) &  0.4299(9)/0.2847(4)  & 0.10386(5)/0.04291(8) \\
$r_{1,2~back}$ &0.1891(7)/0.074(1) & 0.447(1)/0.3171(6)   & 0.10393(5)/0.04294(8)  \\
$\sum W(O-C)^{2}$                    & 0.002 & 0.001 & 0.0005 \\
\hline
Absolute parameters\\
\hline
$M_{1,2}$ ($M_{\odot}$)              & $11.4_{-1.3}^{+1.4}$/$3.1_{-0.4}^{+0.4}$ & $17.3_{-1.9}^{+2.0}$/$6.2_{-0.7}^{+0.7}$ &$12.8_{-0.7}^{+0.7}$/$3.5_{-0.2}^{+0.2}$  \\
$R_{1,2}$ ($R_{\odot}$)              & $5.9_{-0.2}^{+0.2}$/$2.32_{-0.09}^{+0.09}$  & $8.9_{-0.3}^{+0.3}$/$6.1_{-0.2}^{+0.2}$ & $6.3_{-0.1}^{+0.1}$/$2.59_{-0.05}^{+0.05}$ \\
$\log(g_{1,2})$ (cgs)                & $3.95_{-0.02}^{+0.02}$/$4.19_{-0.02}^{+0.02}$ &$3.78_{-0.02}^{+0.02}$/$3.66_{-0.02}^{+0.02}$  & $3.952_{-0.008}^{+0.008}$/$4.159_{-0.008}^{+0.008}$ \\
$\log(L_{1,2}/L_{\odot})$            & $4.13_{-0.04}^{+0.03}$/$2.17_{-0.04}^{+0.04}$ & $4.44_{-0.03}^{+0.03}$/$3.42_{-0.03}^{+0.03}$ & $4.21_{-0.02}^{+0.02}$/$2.55_{-0.02}^{+0.02}$\\
$M_{\rm bol,1,2}$ (mag)              & $-5.58_{-0.09}^{+0.09}$/$-0.68_{-0.09}^{+0.09}$ & $-6.37_{-0.08}^{+0.09}$/$-3.81_{-0.08}^{+0.08}$ &$-5.78_{-0.04}^{+0.04}$/$-1.64_{-0.04}^{+0.04}$  \\
\hline
Other derived parameters\\
\hline
$P_{r1,2}$ (d)                         &1.67(21)/- &2.27(29)/1.45(16) &1.26(015)/- \\
\hline
\end{tabular}
\renewcommand{\arraystretch}{1.0}  % Default is 1.0
\end{center}
\end{table*}

\subsection{Single-lined spectroscopic binaries (SB1s)}\label{sec:SB1}
For SB1 systems, the light and RV curves were fitted independently because they mainly constrain different parameters: the light curve constrains the eclipse geometry, while the single-lined RV data constrain $K_{1}$ and the systemic velocity $\gamma$. Since $K_{2}$ is not available, the RV data do not provide a direct spectroscopic mass ratio to strongly couple the two solutions. The light curves are modelled using  \textsc{JKTEBOP} \citep{Southworthetal2004}. A logarithmic limb darkening law is used and the gravity- and limb-darkening coefficients are fixed to their appropriate values from the Claret gravity- and limb-darkening tables \citep{Claret2017}  and the sum of radii, fractional radii $k$, inclination $i$, surface brightness ratio $j$, $e\cos\omega$, $e\sin\omega$, period and epoch, and sometimes the reflection coefficients are fitted. For systems such as HD~101838, HD~108628 and HD~339003 with large pulsation amplitude to eclipse depth ratio, we partly removed the pulsation signals before fitting the binary models. Moreover, $j$ is fixed for HD~108628 after preliminary fit to the light curve.

Parameter uncertainties were estimated using the Monte Carlo routine (Task 8) implemented in \textsc{JKTEBOP} \citep{Southworthetal2004, Southworthetal2005}. The best-fitting model was repeatedly perturbed with Gaussian noise and refitted, and the resulting parameter distributions were used to derive the $1\sigma$ uncertainties.

For RV and orbital fitting, the \textsc{RVFIT} code is used, which uses the adaptive simulated annealing method or algorithm to fit the radial velocities of binaries and exoplanets \citep{Iglesiasetal2015}. \textsc{RVFIT}, due to its usage of its global minimisation algorithm, has the capacity to rapidly converge to a global solution minimum without the need to provide preliminary parameter values \citep{Iglesiasetal2015}. However, where plausible preliminary parameter values or geometrical information of the system exist in the literature, the fits are obtained by fixing these available parameters. In most systems, the photometric period and epoch obtained from the light curve modelling were fixed during RV fitting. The preliminary eccentricities of the systems were also obtained from the spectra and significance was tested using the Lucy-Sweeney test \citep{LucyandSweeney1971} and those that did not show significance at the 95\% confidence level were adopted as circular orbits and fixed to zero. However, where the light curve models have shown that the orbits are eccentric, we fit eccentricity as a free parameter and adopt the fitted eccentricities even if the Lucy-Sweeney test suggests otherwise. For systems where the light curve models suggest circular orbits and are corroborated by the Lucy-Sweeney test, we assumed the eccentricities to be zero. The final fits were made under the imposed conditions, where the period, epoch, and/or eccentricity are fixed, and other parameters are used as free parameters, until a minimum chi-square is attained. In cases where the fixed photometric parameters do not fit well with the spectroscopic RV, they are gradually refitted as free parameters (one less fixed parameter at a time) until a plausible solution is attained. Figure \ref{fig:SB1_fits} shows the fitted light and radial velocity curves and Table \ref{tab:SB1_paramters} shows the parameters of the SB1 systems.

The residuals in the RV curves are visually examined for intrinsic variability or pulsations. A peak-to-peak scatter in the residuals likely indicates the presence of pulsations \citep{Johnstonetal2021}. However, no line profile variability analysis could be done to probe the pulsations further because of the limited amount of spectra per star. 

Finally, CPD$-45$ 3109 and HD 92741 which are SB1s are not modelled in detail due to data limitations. Their analyses are described in detail in Appendix \ref{appendix:sec_data_limitations}.

\begin{table*}[ht]
\begin{center}
\caption{Parameters of the single-lined spectroscopic binary systems in our sample. "f" denotes fixed parameters.}\label{tab:SB1_paramters}
\begin{tabular}[t]{cccccc}
\hline\hline
Star    & HD 101838 & HD 108628 & HD 254346 & HD 339003 & V4386\, Sgr \\
\hline\hline
Adjusted Quantities\\
$P$ (d)$^{f}$	 & 5.41187& 4.22358 &5.43175 &6.16599 &10.79916\\
$T_p$ (HJD-2457000)$^{f}$	 &1579.16416 & 3070.67632 &2478.47865 &2424.69931 & 1664.16056 \\
$\gamma\, (\rm km\,s^{-1})$	&0.48(4.31) &$-7(3)$ & 21(2)&$-3(3)$ & $-44(1)$\\
$K_1\, (\rm km\,s^{-1})$		 &22(5) &20(4) & 19(3)&28(4) &30(2)  \\
$r_{1}+r_{2}$    &0.365(6) &0.207(5) &0.247(6) &0.287(3) &0.261(9) \\
$k$   &0.1454(8) & 0.146(1)& 0.205(20)&0.2370(7) &1.3(2) \\
$i$ ($^\circ$)  &83.023(1) & 81.8(3)& 78.8(4)&89.9(1.2) & 78.9(8)\\
$j$  & 0.02(6)& 0.0008$^{f}$& 0.095(6)&0.044(5) & 0.159(4)\\
$e\cos\omega$ &-0.034(1) &- &-0.0101(2) &- & -\\
$e\sin\omega$ & 0.005(6)& -& -0.081(6)&- & -\\
$l_{\rm refl,1}$ & 0.0027(1)& 0.00053(4)&0.00042$^{f}$ &0.0002 & -\\
$l_{\rm refl,2}$ & 0.0043(1)&- & 0.00149$^{f}$&0.0066 &- \\
$\ell_{3}$  & -& -& 0.005(10)&- & -\\
$u_{1}^{f}$   &0.3808 &0.3273 & 0.33&0.33 &0.18 \\
$u_{2}^{f}$  & 0.3622&0.4323 &0.34 & 0.34& 0.60\\
$v_{1}^{f}$   &0.2808 &0.2311 & 0.24&0.24 & 0.16\\
$v_{2}^{f}$   &0.2153 &0.2376 &0.206 &0.206 & 0.20\\
$y_{1}^{f}$   & 0.4278& 0.3867&0.4159 &0.4159 &0.32 \\
$y_{2}^{f}$   &0.3529 & 0.4183& 0.3386&0.3386 & 0.32\\
\hline
Derived Parameters\\
$e$			  &0.034(1) &$0^{f}$ & 0.082(6)& $0^{f}$&0 \\
$\omega$ ($^\circ$)	  &172(10) & 90 &262.9(6) &90 &90  \\
$r_{1}$  &0.318(4) & 0.181(4)&0.205(2) &0.232(2) &0.116(7) \\
$r_{2}$  & 0.046(5)& 0.0264(8)&0.042(5) & 0.0549(6)& 0.15(2)\\
light ratio  & 0.0037(5)& 0.0000165(3)&0.0051(8) &0.0024(3) & 0.2199(5)\\
$a_1\sin i$ ($10^6$ km)	 &1.6(4) &1.2(2) &1.4(2) &2.4(3) & 4.5(2) \\
$f(m_1,m_2)$ ($M_\odot$)	  &0.006(4) &0.004(2) &0.004(2) &0.014(6) &0.031(5)  \\
\hline
Other Quantities from RV\\
$\chi^2$		 &6.26 &3.74 &5.78 &6.71 &27.47 \\
$rms_1\, (\rm km\,s^{-1})$  &10.72 &5.89 & 5.79& 7.59& 7.89 \\
%\\
%Table \ref{tab:orbital-element} continues\\
\hline\hline
\end{tabular}
\end{center}
\end{table*}

\section{Absolute Parameters}\label{subsec:absolute-parameter}
The absolute parameters of the reported SB2 systems in this work are derived from the WD code and their uncertainties are obtained by sampling the distribution of the modelled parameters as described in Section \ref{sec:SB2}. The absolute parameters of the SB2 systems are shown in Table \ref{tab:SB2_parameters} and an example corner plot showing their derived uncertainties is shown in Figure \ref{Appendix:fig_corner_plot_V1166}. The derived solutions are approximately $5 - 12\%$ precision in mass and $1.5 - 3.9\%$ precision in radius. Such high uncertainties especially in mass is dominated by the uncertainty in the semi-major axis arising possibly from the weak low quality secondary  RVs or a very small fraction of the total light contributed by the secondary components.

For SB1 systems, it is only possible to directly measure the binary mass function from the spectra. The binary mass function is defined  as: 
 \begin{equation}
    f(M) = \frac{M^{3}_{2}\sin^{3}\,i}{(M_{1}+M_{2})^{2}} =\frac{PK_{1}^{3}}{2\pi G}\,(1-e^{2})^{3/2}
   \end{equation}
\noindent
where $e$ is eccentricity, $i$ is inclination, $K_{1}$ is radial velocity semi-amplitude, $G$ is the gravitational constant and $P$ is the orbital period.
For a circular orbit ($e=0$), the mass function becomes 

 \begin{equation}
      f(M) = \frac{M^{3}_{2}\sin^{3}\,i}{(M_{1}+M_{2})^{2}} =\frac{PK_{1}^{3}}{2\pi G}   
   \end{equation}

If $M_{1}>>M_{2}$, the mass function becomes approximately $\frac{M^{3}_{2}\sin^{3}\,i}{M_{1}^{2}}$. It becomes approximately ${M_{2}\sin^{3}\,i}$ for $M_{1}<<M_{2}$. Since $0 \leq \sin i \leq 1$ for $0 \leq i \leq 90$, the mass function gives the minimum possible mass of the unseen object. 
Without the inclination and at least the mass of one of the components, we are stuck.

To estimate the evolutionary mass of the primary component, we fitted the observed properties of the primary component using 
the \textsc{BONNSAI} Bayesian tool \citep{Schneideretal2014}, adopting 
the luminosity $L$, effective temperature $T_{\rm eff}$, surface gravity 
$\log g$, and projected rotational velocity $v \sin i$ as observational 
constraints, and the \textsc{BONN\_WM} stellar evolutionary models 
\citep{Brottetal2011}. We also adopted a Salpeter initial-mass-function 
prior \citep{Salpeter1955} and the initial rotational-velocity prior 
based on the Galactic and LMC B-star distributions of 
\citet{Hunteretal2008}. The posterior distributions provided by 
\textsc{BONNSAI} were then used to infer the properties of 
the primary component. The $\log g$ and $v \sin i$ add additional constraints on the radius and age of the star, thus, yielding more precise and accurate stellar parameters for the primary.

\begin{figure}[ht]
\centering
\includegraphics[width= \linewidth]{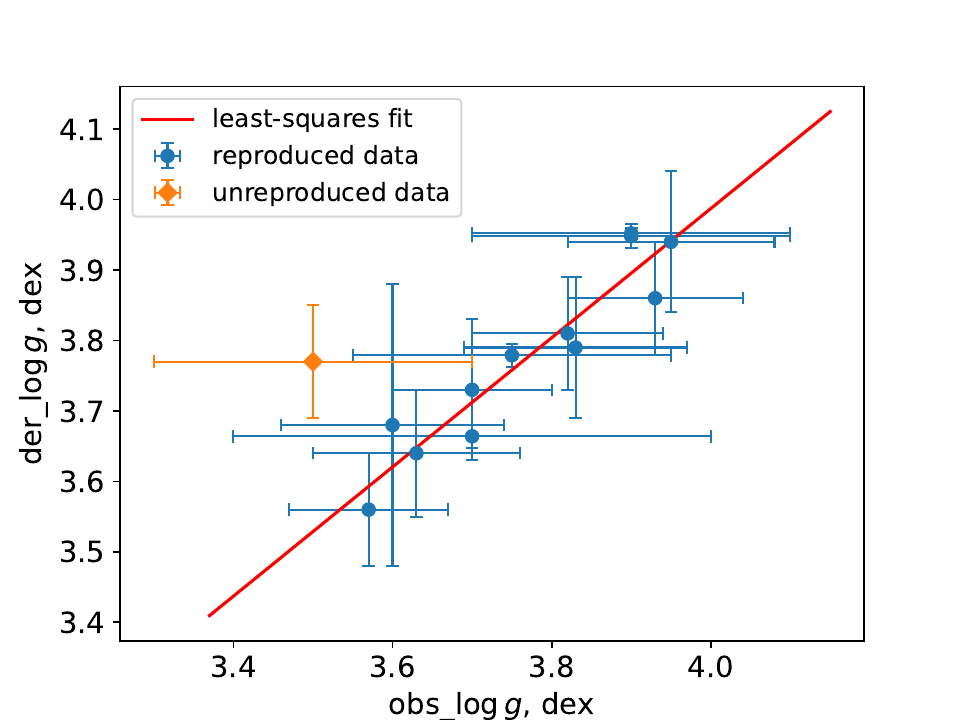}
\caption{Plot of derived $\log g$ against observed $\log g$. The observed  $\log g$ of HD 112026 plotted in orange colour could not be reproduced by the isochrone fitting model, while that of HD 101838 was reproduced although with very large uncertainties in its derived $\log g$. The dynamical $\log g$ derived for the SB2 systems from the binary modelling  are the most precise and also well reproduced. However, the uncertainties in their observed spectroscopic measurements are large.}\label{fig:fig_logg_logg}

\end{figure}

To validate the veracity of the derived parameters, we plotted the $\log g$  derived from the reported parameters against the observed $\log g$ and fitted a linear regression line to it to observe the scatter across the line as shown in Figure \ref{fig:fig_logg_logg}. As seen in the figure, the $\log g$ of HD 112026 could not be reproduced within the observed error. Such is not strange considering the defects in the spectra of HD 112026 that impaired the measurement of the atmospheric parameters from the desired lines. The parameters of CD-38 4128 were fitted without the luminosity as an input parameter due to the very large error observed in the luminosity of the system. The secondary masses are derived, using $M_{1}$ and $i$ shown in this work for the systems, by numerically solving the exact spectroscopic mass-function equation ($(M_2 \sin i)^3 - f(M)\,(M_1 + M_2)^2 = 0$). Uncertainties were estimated using Monte-Carlo sampling of the mass function, primary mass, and inclination. 

The absolute radius of the secondary component was derived from the combination of the light curve and spectroscopic orbital solution using $R_2 = r_2 a$, where $a$ is the semi-major axis, determined from the Kepler's third law as:    
\begin{equation}
a = \left[\frac{G(M_1+M_2)P^2}{4\pi^2}\right]^{1/3},
\end{equation}

\noindent
where $P$ is the orbital period, $M_1$ and $M_2$ are the masses of the primary and secondary components, respectively, and $G$ is the gravitational constant. The fractional radius of the secondary $r_{2}$ was calculated from the fitted sum of the fractional radii, $s=r_1+r_2$, and the radius ratio, $k=r_2/r_1$, according to

\begin{equation}
r_2 = \frac{sk}{1+k}.
\end{equation}

The uncertainties in the secondary radius were determined using Monte Carlo error propagation. The input parameters were randomly sampled within their measured uncertainties, and the median and $1~\sigma$ significance (i.e. 68\% credible interval) of the resulting radius distribution were adopted as the best estimate and its $1\sigma$ confidence interval. 

As an internal consistency check, the spectroscopic mass function was calculated independently from the radial-velocity parameters ($P$, $e$, and $K_{1}$ as well as  from the component masses and orbital inclination, and compared with the observed mass function from the orbital solutions. They all agree within $1\sigma$.

\begin{figure*}[ht]
\centering
\includegraphics[width=\linewidth]{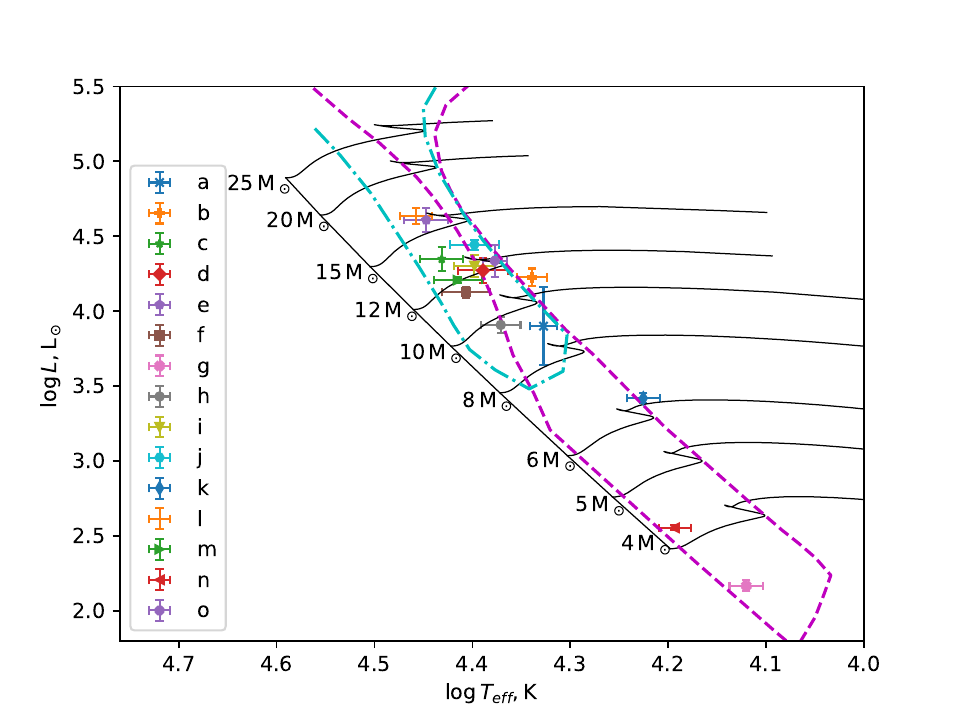}
\caption{HR diagram of the sample showing $\beta$ Cep and SPB instability strips. Note: (a) CD$-38$ 4128, (b) CPD$-45$ 3109, (c) HD 101838, (d) HD 108628, (e) HD 112026, (f) HD 112485A, (g) HD 112485B, (h) HD 157400, (i) HD 254346, (j) HD 329379A, (k) HD 329379B, (l) HD 339003, (m) V1166 Cen~A, (n) V1166 Cen~B, (o) V4386 Sgr. }\label{fig:HR_diagram}

\end{figure*}

The luminosity, $L$ used in the fitting of isochrones and the plotting of the HR diagram as shown in Figure \ref{fig:HR_diagram}, with $\log (L/L_\odot)$ as the ordinate and  $T_{\rm eff}$ as the abscissa, is calculated using the equation $\log (L/L_\odot) = 0.4\times(4.74-M-BC)$ in the case of SB1. M, which is the absolute magnitude, is defined as $M = m-$5$\times \log$(1000/$\pi$)$+$5$-$$A_0$ and $BC$ is the bolometric correction, where $m$ is the apparent magnitude in the V band, $\pi$ is the parallax and $A_0$ is the interstellar extinction. $\pi$ is obtained from the Gaia DR3 catalogue \citep{Gaiacollaborationetal2016b,Gaiacollaborationetal2023j} and $T_{\rm eff}$ is the spectroscopic $T_{\rm eff}$ reported for the systems in this work and shown in Table \ref{tab:atmospheric-solutions}. $A_0$ is derived using the information available in \citet{PecautandMamajek2013} and Simbad. We computed the expected (B-V) colour index from standard relations \citep{PecautandMamajek2013}, obtained the observed (B-V) from Simbad and multiplied the difference between these two values, which we assume to be the colour excesses with 3.2, which is the current value of $A_{V}/E(B-V)$. The bolometric correction and its uncertainties are calculated in line with \cite{1996ApJ...469..355F}. The errors in $L$ are derived using standard error propagation. The theoretical evolutionary tracks used are the precomputed tracks for masses 4, 5, 6, 8, 10, 12, 15, 20 and 25$M_\odot$ with the Warsaw-New Jersey evolution and pulsation code (described, for instance, by \citealt{1998A&A...333..141P}) using $Z = 0.012$, $X = 0.700$ and {$ \sin i_{ZAMS}=100\,\rm km\,s^{-1}$}. Theoretical instability strips are also assigned using the same conditions \citep{1999AcA....49..119P}. Figure \ref{fig:HR_diagram} gives the HR diagram, and Table \ref{tab:other_derived_parameters_for_SB1} shows the derived absolute stellar parameters of the  SB1 systems in the sample. We note that although the nominal position of CPD$-$45 3109 lies outside the $\beta$ Cep instability strip, it is consistent with the edge of the strip within uncertainties of its parameters for the adopted model metallicity. 

%------------------------------------------------------------------------%-----

%------------------------------------------------------------------------
\section{Circularisation and Synchronisation}\label{subsec:circularisation-and-synchronisation}
Close binary systems experience tidal forces which affect both the shape of the orbits as well as the rotational synchronisation of the stars. This results in tidal circularisation of the orbit or tidal synchronisation of the rotation of the system \citep{Zahn1977, Hut1981}. For tidal circularisation, eccentric orbits gradually lose their eccentricity and become increasingly circular over time owing to energy dissipation by tidal forces \citep{GoldreichandSoter1966, TerquemandMartin2021}, and for tidal synchronisation, the tidal interaction between stars creates forces that gradually change their rotation rates and align them with the orbital motion such that the rotational period becomes equal to the orbital period of the system \citep{Zahn1977, Hut1981}.

\begin{table*}[ht]
\begin{center}
\caption{Derived stellar parameters for the SB1 systems.  }\label{tab:other_derived_parameters_for_SB1}
\begin{tabular}[t]{lcccrrrr}
\hline\hline
Star & $M_{1} (M_\odot)$   & $R_{1} (R_\odot)$&  Age (Myr) & $M_{2} (M_\odot)$ & $R_{2} (R_\odot)$& $\log L (L_\odot)$ &	$P_{r}$ (d) \\
\hline\hline
CD$-38$ 4128	&	8.8(8)	&	6(1)	&	21.2(2.8)	&		-&		-&3.9   (2)	  & -\\
CPD$-45$ 3109		&	11.0(4)	&	9.0(8)	&	16.6(1.2)	&	1.3(3)&		-&	4.23(6)&2.9(3)  \\
HD 101838		&	13.2(8)	&	7.4(9)	&	10.2(1.7)	&	1.1(3)	&	1.46(3)	&	4.35(8)& 1.9(3)\\
HD 108628		&	12.0(7)	&	6.7(6)	&	10.8(1.8)	&	0.9(2)	&		0.68(2)&4.27(8)	&2.4(4) \\
%HD112026*	&	15.4(9)	&	9(1)	&	8.6(1.1)	&	3.4(4)	&		&	4.61(8)& 7(1)\\
%HD 157400	&	9.8(4)	&	5.4(5)	&	13.8(3.4)	&	3.8(6)	&		2.60(5)&3.91(5)	& 1.4(2)\\
HD 254346	&	12.2(6)	&	7.3(8)	&	11.5(1.7)	&	0.9(2)	&	1.29(8)	&	4.30(7)&2.1(3) \\
HD 339003		&	16.4(7)	&	8.3(8)	&	7.2(9)	&	1.7(3)	&	2.04(3)	&4.64(5)	& 1.9(2)\\
V4386 Sgr	&	12.0(8)	&	8(1)	&	13.2(1.1)	&	1.8(1)	&	7.1(5)	&	4.3(1)&5.1(9) \\
\hline\hline
\end{tabular}
\end{center}
\end{table*}

\begin{figure}[ht]
\centering
\includegraphics[width=\linewidth]{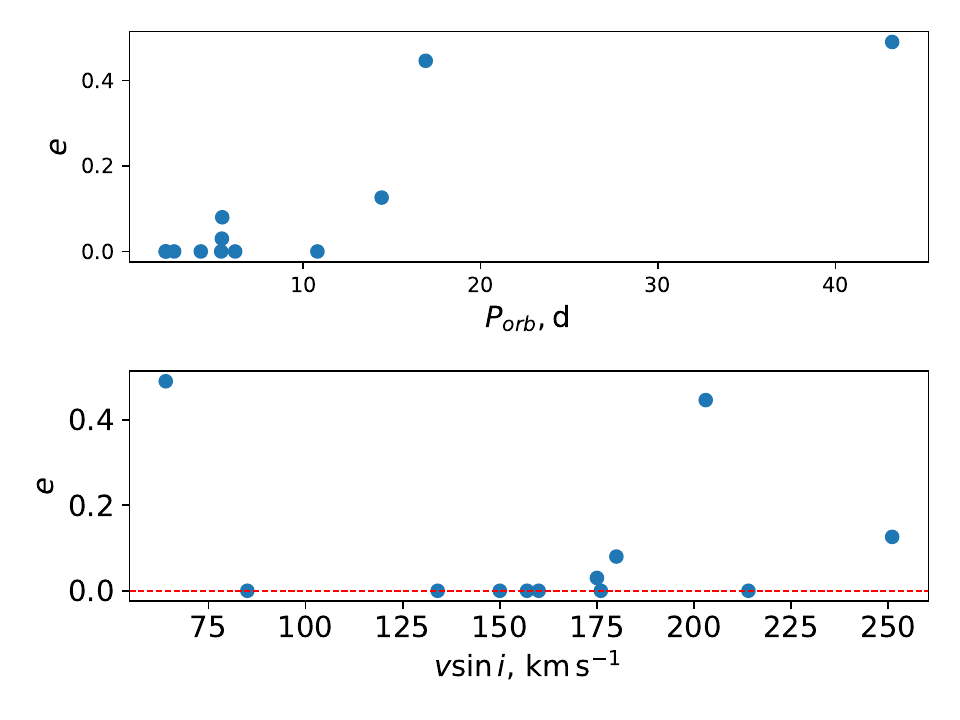}
\caption{The dependence of eccentricity on orbital period as well as projected rotational velocity.}\label{fig:e_p_v_relation}

\end{figure}

 Several authors have investigated circularisation and synchronisation. \citet{MeibomandMathieu2005} studied solar-type stars and determined an eccentricity-orbital period relation. They observed a sharp change from eccentric to circular orbits at a period of approximately 10 days. This finding was corroborated by \citet{Lurieetal2017} and \citet{Mirouhetal2024} for stars of masses $0.7 - 1.6\,M_\odot$. Eclipsing binaries involving Kepler targets are also found to be synchronised at orbital periods of less than 10 days \citep{Lurieetal2017}. For O stars, the transition from eccentric to circular orbits occurs at a period of approximately 6 days \citep{Mirouhetal2024}, as they evolve and circularise faster than their low mass counterparts. Circularisation timescales depend strongly on binary separation, evolutionary stage, and stellar structure \citep{Zahn1977} and highly eccentric orbits are also observed to circularise much faster than moderately eccentric orbits \citep{TerquemandMartin2021}. 
 To investigate the circularisation of the systems in our sample, we plotted their eccentricity against some stellar parameters such as the orbital period and projected rotational velocity, as shown in Figure \ref{fig:e_p_v_relation}. Figure \ref{fig:e_p_v_relation} (top panel) shows that the orbital period transition window from eccentric to circular orbit is $5\leq P\leq20$ for the current sample, which is consistent with the literature \citep[e.g.][]{Mirouhetal2024}. Below this range, the systems are purely circular and above it, the systems are purely eccentric. The systems in our sample are predominantly circular. This is understandable, as OB binaries tend to have more circular orbits \citep{VargasSalazaretal2025}. Short-period orbits are also easier to detect, keeping in mind that our sample was constructed from photometry of predominantly 27-d light curves. 
 
 Eccentricity appears to have a more complicated relationship with the projected rotational velocity ($v \sin i$). The evidence available so far in the literature appears to suggest a system-specific relationship of the parameters rather than following a universal pattern. An eccentricity of 0.398 was found, for instance, by \citet{Putkurietal2018} for a binary system ($P=50.432(1)\,\rm d$) of two O-type stars with projected rotational velocities of 65 and 325 $\,\rm km\,s^{-1}$ for the components. Several factors, such as stellar evolution, core rotation rates, and possible misalignment between the rotational and orbital axes, significantly influence the evolution of the eccentricity and possibly how it relates to other parameters \citep{Becketal2014}. However, for systems with considerably similar properties, the eccentricity could show some trends.  In Be binaries, for instance, a high eccentricity at high $v \sin i$ is likely observable due to the kicks by the supernova remnants \citep{Martinetal2009}. In the $\beta$ Cep sample in this work, no significant correlation between $e$ and $v \sin i$ is observed. Nevertheless, certain trends could still be observed if several $v \sin i$ regimes are considered separately. The systems appear to be predominantly eccentric for $v \sin i < 75 \,\rm km\,s^{-1}$ and $v \sin i > 175 \,\rm km\,s^{-1}$. For systems with $75 \leq v \sin i \geq 175 \,\rm km\,s^{-1}$, the eccentricity approximates zero, implying a circular configuration.  More data is needed to properly characterise the $e$-$v \sin i$ relationship in OB binaries.

\begin{figure}[ht]
\centering
\includegraphics[width=0.95\linewidth]{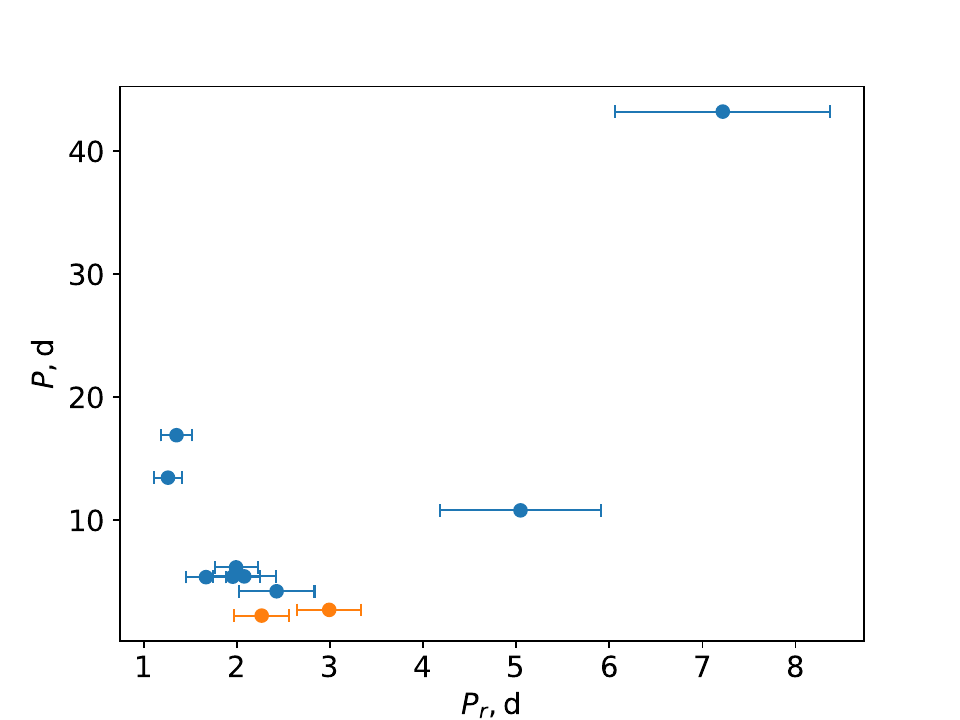}
\caption{The plot of the orbital period against the rotational period. The blue colour shows non-synchronised systems and the orange colour shows synchronised systems. }\label{fig:P_Prot_plot}
\end{figure}

For a synchronised system, the stellar rotational period $P_{r}$ is equal to the orbital period $P$. To check for synchronisation, we first estimate the rotational period using the relation $P_{r} = \frac{2\pi R\sin i}{v \sin i}$, where $v \sin i$ and $R$ are the projected rotational velocity and the radius. The orbital inclination reported for the systems in this work is adopted as $i$. This assumes alignment between the stellar rotational and orbital axes, $i_{\rm rot}=i_{\rm orb}$. In the absence of independent constraints on the stellar spin inclinations, spin–orbit misalignment cannot be excluded. If $i_{rot} \neq i_{orb}$, the inferred rotational period is modified by a factor $\frac{\sin i{\rm rot}}{\sin i_{\rm orb}}$, introducing an additional systematic uncertainty into the assessment of rotational synchronisation. The resulting synchronisation classifications should therefore be interpreted subject to this assumption. A rough agreement between the orbital period and the estimated rotational period is reported as possible synchronisation of the systems. The rotational periods  of the SB2 and SB1 systems with sufficient data are captured in Tables \ref{tab:SB2_parameters} and \ref{tab:SB1_paramters}, respectively.  Figure \ref{fig:P_Prot_plot} shows the plot of the orbital period against the rotational period for the targets in the sample. Two systems are observed to be synchronised within their observed uncertainties. 

As clearly visible in Figure \ref{fig:P_Prot_plot}, there are non-synchronised close binary systems, whose orbital periods show no particular trend with the rotational periods. The coupling of gravity with angular momentum transport determines the surface rotation periods of stars. In synchronised systems, the tides have sufficiently coupled them so that the orbital period of the system is equal to the rotational period of the star, resulting in an already established linear trend in the literature for circularised and synchronised short-period systems. However, this is not the case for non-synchronised short-period systems, whose behaviour appears to depart from what is expected. Circularised systems with $P\leq 10\,\rm d$ are reported to be synchronised, at least in the low mass regime \citep{Lurieetal2017}. As a result, we expected circularised close massive binary systems to be synchronised, but that was not completely the case as some are but some others are not.  
This departure possibly suggests that for most massive close binary systems, the circularisation and synchronisation timescales differ, and may be impacted by the interplay between tides, winds and angular momentum transport. For the eccentric wide binaries with long orbital periods, the data is not sufficient to search for a possible trend. More data are needed to investigate any possible statistical relationship between P and $P_{r}$.

%-------------------------------------------------------------------------
\section{Discussion}\label{sec:discussion}
%-----------------------------------------------------------------------
In this paper, thirteen eclipsing binary systems are analysed. Detailed stellar parameters are reported for three SB2 and five SB1, the parameters partly reported for five other systems.  
The periods reported in this work agree considerably with the literature \citep[e.g.][]{EzeandHandler2024a}, except for HD 157400 with a period twice the value reported in \citet{EzeandHandler2024a}. 

The three SB2 systems are newly identified in this work and provide direct constraints on the fundamental properties of their stellar components. The dynamically determined masses of the primary components span approximately $11$ -- $17~\rm M_{\odot}$, placing them in the massive B-star regime, whereas the secondary components have considerably lower masses of approximately $3$ -- $6~\rm M_{\odot}$, placing them in the intermediate mass regime. The precision of the measured dynamical masses is approximately $5$ -- $12\%$, with the uncertainties dominated by uncertainties in the semi-major axes, which are likely caused by the low flux contributions by the secondary components. Radii are measured more precisely with $1.5$ -- $3.9\%$ precision. In addition, the high luminosity/flux contrast between the components explains why the secondary spectra are very weak or barely detectable in the SB2 systems. Also, HD~112485 and HD~329379 do not have third light contributions, while the inferred third light contribution for V1166 Cen is statistically consistent with zero. 

This study constitutes one of the largest sample of $\beta$ Cep eclipsing binaries with detailed light and radial velocity modelling of the targets. It has added three SB2 systems with dynamically derived parameters  and no less than five SB1 systems  to the pool of $\beta$ Cep eclipsing binaries with detailed reported orbital solutions and absolute stellar parameters.

Here, HD 112485 was reported as SB2 contrary to its treatment as SB1 in \citet{Nazeetal2025}. However, the parameters of its primary components agree with the findings in \citet{Nazeetal2025}. The orbital elements recently reported for V1166 Cen by \citet{Ezeetal2025} were revised here using disentangling. As a result, $K_{1}$ and $K_{2}$  for the components were derived instead of only $K_{1}$ that was previously reported. The $K_{1}= 49.5(9)~\rm km\,s^{-1}$ agree with the previously reported value of $K_{1}= 48(5)~\rm km\,s^{-1}$.

Spectroscopy provides direct line-of-sight measurements that reflect the orbital dynamics in binaries and is devoid of the complexities of "wobbles" in proper motions. Before this work, \citet{DrobekandPigulski2014} had reported the spectroscopic radial velocity semi-amplitude, $K_{1} = 31.8 \pm 0.6\,\rm km\,s^{-1}$ for V4386 Sgr, which is corroborated by the findings in this work. V4386 Sgr has a low-mass secondary component with a larger radius and Roche-lobe filling fraction than its primary component.

The properties of HD 254346 were also recently reported by \citet{Cakirlietal2025} and \citet{Nazeetal2025}. However, their results show discrepancies with each other and partly with the findings in this work.  Again, a mean radial velocity measurement of $10(2)~\rm km\,s^{-1}$ was obtained for CD$-38$ 4128, but there was no plausible radial velocity fit. As a result, its radial velocity fit or semi-amplitude is not reported here. However, its radial velocity measurements are shown in Table \ref{appendix:measured-rvs}. The mean radial velocity of $16.6(1)~\rm km\,s^{-1}$ was also reported for HD 339003 by \citet{Jonssonetal2020}.

We report circular orbits in seven systems, eccentric orbits in five and no orbital solutions in one system. It has been reported in the literature \citep[e.g.][]{TerquemandMartin2021} that eccentric orbits circularise over time owing to tides and that the circularization timescale depends on the structure and evolutionary state of the system. The period threshold for circularisation has been found to be $P<10\,\rm d$ for solar-type stars and $P < 6\,\rm d$ for O stars of masses $20$ -- $80\, M_\odot$ \citep{Mirouhetal2024}. Our sample, which contains B-type stars of masses $9$ -- $17\,\rm M_\odot$, also shows a period window of $5\leq P_{\mathrm{orb}}\leq20~\rm d$  for circularisation, in agreement with the findings in the literature for O stars. This implies that generally, for massive stars of O and B types, their circularisation follows a similar timescale. 
Again, as earlier stated, \citet{Lurieetal2017} reports that low-mass eclipsing binaries with $P_{orb} < 10\,\rm d $ are also synchronised. However, such findings are not fully corroborated in this work for massive stars as some circularised short-period systems show no synchronisation.  

The systems are slow to fast rotators spanning a range of $v \sin i$ of $15$ -- $252\, \rm km\,s^{-1}$. CD$-38$ 4128 is the slowest, and V1166 Cen is the fastest. HD 339003 also has a high $ \sin i$ of $214 (12)\, \rm km\,s^{-1}$.  The high $v \sin i$ is one of the observational signatures of fast surface rotation in binary stars. During the main sequence lifetime of massive stars, the envelope rotation slows down and the core rotation increases. Depending on the angular momentum transport efficiency, the envelope rotation may be spun up resulting in a higher surface rotation, which impacts $ \sin i$. This invariably makes $v \sin i$ an indirect signature for the interior rotation. Fast internal rotation in massive binary stars is caused by initial angular momentum during formation. Conservation of angular momentum, during the collapse of large molecular clouds from which massive stars are formed, leads to rapid rotation. In addition, binary interaction resulting in mass and angular momentum transfer and internal magnetic dynamics \citep[e.g][]{Nazeetal2023, Hirschietal2024} also drive fast rotation in massive stars. On the other hand, fast rotation also has other diverse effects or impacts on the evolution and dynamics of massive stars, among which are splitting or modification of internal pulsation modes \citep[e.g.][]{Bugnetetal2021, Guoetal2024}, enhanced mass loss through stellar wind \citep[e.g.][]{PerezRendonetal2011, Hirschietal2024}, enhanced internal mixing \citep[e.g.][]{PerezRendonetal2011, Hirschietal2024} and mass transfer and spin up in interacting binaries \citep[e.g.][]{deMinketal2013}. Although these impacts are beyond the scope of this work and will not be discussed individually, they, however,  portray stars in our sample, which are fast-rotating B-type pulsating stars in eclipsing binaries, as veritable testbeds for different astrophysical processes or stellar evolution scenarios in massive stars. The orbits of the systems also appear to be affected by the magnitude of $v \sin i$ as reported in this work. The orbits are eccentric for $v \sin i < 75~\rm km\,s^{-1}$ and circular for moderately fast rotators ($75 \leq v \sin i \leq 175$) and predominantly eccentric  extreme cases of $v \sin i$.

%------------------------------------------------------------------------------------------
\section{Conclusions}\label{sec:conclusion}
A sample of 13 $\beta$ Cephei stars in eclipsing binaries is spectroscopically and photometrically analysed to obtain their orbital and absolute stellar properties and infer the relationship between orbital and stellar parameters. Prior to this work, there were a few works on $\beta$ Cep stars in eclipsing binaries with orbital solutions and stellar parameters available in the literature (see Section \ref{sec:introduction}). This paper significantly increases the size of massive pulsating eclipsing binaries with orbital and stellar parameters, paving way for in-depth and ensemble asteroseismology.  

The systems analysed in this work are predominantly fast asynchronous rotators with  seven showing circular orbits and five showing eccentric orbits. No orbital solution was found for one system. The surface rotation of two systems appear to be synchronised with their orbital rotation. The estimated effective temperatures and masses of the primary components of the systems range from $22000$ -- $28000\,\rm K$ and $9$ -- $17\,\rm M_\odot$,  respectively. The uncertainties in the masses and radii limit the precision of the dynamical characterisation of the systems, particularly for the masses, for which the uncertainties reach $5$ -- $12\%$. The secondary components appear to have masses ranging from low-mass to intermediate mass stars.  The eccentricity of the systems shows no universal correlation with $v \sin i$, but shows several trends at different $v \sin i$ regimes.

\begin{acknowledgements}
This work was supported by the Polish National Science Foundation (NCN) under grant nr. 2021/43/B/ST9/02972. CIE thanks Dr. Ahmet Dervisoglu for useful inputs.
Some spectra of the observations reported in this paper were obtained with the Southern African Large Telescope (SALT). Polish participation in SALT is funded by grant No. MEiN nr 2021/WK/01. We also acknowledge the use of the HERMES spectrograph on the Mercator telescope, CHIRON spectrograph operated by the SMARTS Consortium, and the Skalnaté Pleso Observatory, managed by the Astronomical Institute of the Slovak Academy of Sciences for these observations. This paper also used the TESS data obtained from the Mikulski Archive for Space Telescopes (MAST) at the Space Telescope Science Institute. Support to MAST for these data is provided by the NASA Office of Space Science via grant NAG5-7584 and by other grants and contracts. Funding for the TESS mission is provided by the NASA Explorer Program.
The paper used the SIMBAD database and the VizieR catalogue access tool, operated at CDS, Strasbourg, France; and the SAO/NASA Astrophysics Data System. It also used data from the European Space Agency (ESA) mission
{\it Gaia} (\url{https://www.cosmos.esa.int/gaia}), processed by the {\it Gaia}
Data Processing and Analysis Consortium (DPAC;
\url{https://www.cosmos.esa.int/web/gaia/dpac/consortium}). Funding for the DPAC
has been provided by national institutions, in particular the institutions
participating in the {\it Gaia} Multilateral Agreement.

\end{acknowledgements}

%--------------------------------
 \bibliographystyle{aa} % style aa.bst
\bibliography{bibliography} % your references Yourfile.bib

%---------------------------------------------
\begin{appendix}
\section {Orbital fitted parameters}\label{appendix:orbital_fitted_parameters}

The stated RVs here are derived in line with Section \ref{subsec:orbital-variability}, except those of HD 112485, HD 329379 and V1166 Cen which are derived in line with section \ref{sec:SB2}
\newline
\newline

\setlength{\tabcolsep}{1.8pt}
\renewcommand{\arraystretch}{0.97}

\topcaption{Measured radial velocities of the stars rounded off to the nearest precision.}
\label{appendix:measured-rvs}

\begin{supertabular}{@{}lcccc@{}}
Target & HJD & $RV_1$ & $RV_2$ & $\phi_{\rm orb}$ \\
       &     & $(\mathrm{km\,s^{-1}})$ & $(\mathrm{km\,s^{-1}})$ & \\
\hline
\bf{1. CD$-$38 4128}& &- &  \\
& 2458799.50179	& 12(1)	& -&0.150  \\
& 2458802.50165	 & 15(1)	& -& 0.673\\
& 2458803.49680	  & 13(1)	& -&0.847\\
& 2458815.47119	 & 11(1)	& -& 0.937 \\
& 2458829.43794	 & 8(2)	&- & 0.374\\
& 2458830.42292	  & 1(2)	& -&0.546\\
& 2458833.41248	& 3(2)	& -& 0.068 \\
& 2458840.38899	 & 15(1)	&- & 0.285\\
& 2458850.37492	 & 3(2)	&- & 0.028\\
& 2458856.35307	 & 18(1)	& -& 0.071\\
\hline
\bf{2. CPD$-$45 3109}  & & & \\
& 2459916.49046  &    51(6)&	-& 0.891\\
& 2460015.45883   &   $-8(7)$&-& 0.358\\
& 2460016.46528  &    58(7)&	-& 0.729\\
& 2460017.46224  &    19(5)&	-& 0.096\\
& 2460023.44300  &    $-5(5)$&	-& 0.300\\
& 2460025.44464  &    36(8)& -& 0.038\\
& 2460028.43883   &   24(8)& -& 0.141\\
& 2460031.42244   &   $-2(4)$&- & 0.240\\
& 2460032.42405  &    40(7)	&-& 0.610\\
\hline
\bf{3. HD 101838}&  & &  \\
& 2458535.38566 &	$-9(15)$	&-& 0.923\\
& 2458623.20686 & 	$-13(19)$	&-& 0.150\\
& 2458931.31138 &	$-19(21)$	&-& 0.081\\
& 2458980.42266 &	$-20(12)$	&-& 0.156\\
& 2458981.38781 &	$-28(13)$	&-& 0.334\\
& 2458989.22025 &	19(13)	&-& 0.782\\
& 2459001.31958 &	6(10)	&-& 0.017\\
& 2459003.37014 &	$-7(12)$	&-& 0.396\\
& 2459005.36229 &	26(9)	&-& 0.764\\
& 2459006.33076 &	13(11)	&-& 0.943\\
& 2459007.34379 &	$-8(22)$	&-& 0.130\\
\hline
\bf{4. HD 108628}&  & & \\
& 2458985.38968 &	4(8)	&-& 0.656\\
& 2458986.44514 &	$-2(10)$	&-& 0.906\\
& 2459003.37904 &	$-4(9)$	&-& 0.917\\
& 2459005.37167 &	$-19(11)$	&-& 0.389\\
& 2459006.35273 &	10(11)	&-& 0.621\\
& 2459007.29067 &	16(10)	&-& 0.843\\
& 2459008.31391 &	$-8(13)$ &-& 0.086\\
& 2459009.34721 &	$-24(9)$	&-& 0.331\\
& 2459010.38199 &	$-0.3(9.6)$	&-& 0.576\\
& 2459015.31199 &	18(9)	&-& 0.743\\
\hline
\bf{5. HD 112026}&  & & \\
&	2459971.83412	&	$-6.07$	&	21.68	&	0.241\\
&	2459973.8413	&	10.89	&	$-38.92$	&	0.287	\\
&	2459984.81333	&	19.53	&	$-69.76$	&	0.541	\\
&	2459989.80657	&	14.49	&	$-51.75$	&	0.657	\\
&	2459990.80789	&	13.24	&	$-47.29$	&	0.679	\\
&	2459996.76478	&	3.94	&	$-14.06$	&	0.818	\\
&	2460001.79555	&	$-7.79$	&	27.83	&	0.934\\
&	2460007.76306	&	$-32.06$	&	114.49	&	0.072	\\
&	2460012.80558	&	$-37.01$	&	132.17	&	0.189\\
&	2460022.69655	&	21.81	&	$-77.89$	&	0.418	\\
\hline
\bf{6. HD 112485}&  & &   \\
&	2460015.42902	&	22.79	&	$-89.79$	&	0.539\\
&	2460016.57029	&	58.63	&	$-230.96$	&	0.752\\
&	2460017.57615	&	13.9	&	$-54.77$ &	0.939\\
&	2460017.60399	&	12.02	&	$-47.34$	&	0.945\\
&	2460018.43434	&	$-41.21$	&	162.36	&	0.099\\
&	2460018.44639	&	$-41.81$	&	164.72	&	0.102\\
&	2460018.56927	&	$-47.42$	&	186.79	&	0.124\\
&	2460020.57674	&	7.78	&	$-30.64$	&	0.498\\
&	2460022.40854	&	45.29	&	$-178.42$	&	0.839\\
&	2460023.42622	&	$-18.81$	&	74.11	&	0.029\\
&	2460025.55719	&	$-19.06$	&	75.09	&	0.425\\
&	2460031.34952	&	9.72	&	$-38.29$	&	0.503\\
&	2460032.33609	&	57.49	&	$-226.46$	&	0.687\\

\hline
\bf{7. HD 157400} &  & & \\
&	2459701.42251	&	$-32.6$&	256.45	&	0.351\\
&	2459722.61236	&	$-28.03$	&	220.46	&	0.604\\
&	2459723.63176	&	$-22.19$	&	174.59	&	0.665	\\
&	2459750.29076	&	$-22.55$	&	177.41	&	0.242	\\
&	2459815.36869	&	28.93	&	$-227.59$ &	0.091	\\
&	2459846.28804	&	57.02	&	$-448.53$	&	0.920	\\
&	2459849.27928	&	25.06	&	$-197.17$	&	0.097	\\
\hline
\bf{8. HD 254346}&  & & \\
& 2458448.64649	& 9(8) 	&-& 0.405\\
& 2458452.67126	& 5(7)	&-& 0.146\\
& 2458455.61877	& 36(6)	&-& 0.689\\
& 2458487.60643	& 32(6)	&-& 0.578\\
& 2458491.57962	& 6(7)	&-& 0.309\\
& 2458492.53994	& 20(7)	&-& 0.486\\
& 2458493.46961	& 41(8)	&-& 0.657\\
& 2458494.52399	& 35(7)	&-& 0.852\\
& 2458495.52359	& 19(6)	&-& 0.036\\
& 2458496.5023	& 6(7)	&-& 0.216\\
& 2458757.74647	& $-5(8)$	&-& 0.311\\
& 2458758.76245	& 12(9)	&-& 0.498\\
& 2458760.7616	& 24(13)&-& 0.866\\
& 2458762.7411	& $-4(7)$	&-& 0.231\\
& 2460005.69462	& 26(11)&-& 0.061\\
\hline
\bf{9. HD 329379}&  & &   \\
&	2459701.40963	&	14.88	&	$-41.72$	&	0.499	\\
&	2459722.37064	&	70.95	&	$-198.93$	&	0.829	\\
&	2459722.5985	&	23.67	&	$-66.36$	&	0.931	\\
&	2459723.59398	&	$-52.71$	&	147.77	&	0.374	\\
&	2459735.34208	&	66.07	&	$-185.23$	&	0.603	\\
&	2459738.31047	&	27.02	&	$-75.74$	&	0.925	\\
&	2459740.57266	&	23.21	&	$-65.06$	&	0.932	\\
&	2459819.35141	&	$-15.74$	&	44.13	&	0.999	\\
\hline
\bf{10. HD 339003}&  & & \\
& 2458197.74504	& 14(12)	&-& 0.940\\
& 2458200.69709	& $-12(13)$ &	-& 0.419\\
& 2458203.70289	& 5(12)	 & -  & 0.907\\
& 2458316.64031	& $-40(10)$ &	-& 0.229\\
& 2458317.58275	& $-9(12)$	&-& 0.382\\
& 2458318.65104	& 10(11)	&-& 0.555\\
& 2458319.62013	& 13(18)	&-& 0.712\\
& 2458320.57181	& 21(13)	&-& 0.867\\
& 2458321.58559	& $-9(10)$	&-& 0.031\\
& 2458351.45599	& 27(10)	&-& 0.877\\
& 2458372.36121	& $-27(10)$	&-& 0.269\\
& 2458372.37568	& $-28(9)$	&-& 0.271\\
& 2458372.39015	& $-28(10)$	&-& 0.273\\
& 2458603.71638	& 14(11)	&-& 0.802\\
& 2458605.72539	& $-25(9)$	&-& 0.128\\
& 2460005.68909	& $-42(14)$	&-& 0.247\\
& 2460006.16778	& $-38(13)$	&-& 0.325\\
\hline
\bf{11. HD 92741}&  & & \\
& 2459952.81882 &	$-13(11)$	&-& 0.536\\
& 2459954.787820 &	$-2(12)$	&-& 0.903\\
& 2459957.76329 &	$-24(12)$	&-& 0.456\\
& 2459971.74898 &	$-28(11)$	&-& 0.059\\
& 2459972.76342 &	$-54(13)$ &-& 0.248\\
& 2459973.65653 &	$-32(11)$ &	-& 0.414\\
& 2459975.76493 &	16(12)	&-& 0.807\\
& 2459982.72432 & $-36(12)$	&-& 0.102\\
& 2459983.73913 &	$-47(14)$	&-& 0.291\\
& 2459986.75594 &	11(11)	&-& 0.852\\
\hline
\bf{12. V1166 Cen}& &  &\\
&	2459947.5814	&	16.38	&	$-57.89$	&	0.513	\\
&	2459967.59718	&	$-13.94$	&	49.27	&	0.004	\\
&	2459968.59642	&	$-34.62$	&	122.36	&	0.079	\\
&	2459992.81253	&	25.59	&	$-90.44$	&	0.883	\\
&	2459994.83143	&	$-21.29$	&	75.25	&	0.034	\\
&	2459996.77737	&	$-48.66$	&	171.99	&	0.179	\\
&	2459997.61836	&	$-48.43$	&	171.17	&	0.241	\\
&	2459998.78474	&	$-36.42$	&	128.69	&	0.328	\\
&	2460001.80833	&	24.43	&	$-86.33$	&	0.554	\\
&	2460005.78709	&	34.62	&	$-122.36$	&	0.850	\\
&	2460006.58634	&	18.82	&	$-66.51$	&	0.909	\\
&	2460007.77568	&	$-9.51$	&	33.62	&	0.998	\\
&	2460011.77487	&	$-42.65$	&	150.74	&	0.296	\\
&	2460013.81617	&	$-6.07$	&	21.47	&	0.448	\\
&	2460015.44439	&	27.74	&	$-98.03$	&	0.569	\\
&	2460016.47556	&	42.98	&	$-151.9$	&	0.647	\\
&	2460016.56272	&	43.87	&	$-155.05$	&	0.653	\\
&	2460017.52969	&	49.14	&	$-173.684$	&	0.725	\\
&	2460017.72899	&	49.04	&	$-173.31$	&	0.739	\\
&	2460018.64118	&	43.16	&	$-152.53$	&	0.808	\\
&	2460020.59238	&	5.99	&	$-21.17$	&	0.953	\\
&	2460020.63728	&	4.91	&	$-17.34$	&	0.957	\\
\hline
\bf{13. V4386 Sgr} & & & \\
& 2458225.51047	& $-69(7)$ & -& 0.373\\
& 2458237.47554 & $-58(22)$ &-& 0.481\\
& 2458985.44277	& $-19(11)$ &-& 0.750\\
& 2458989.66824	& $-77(6)$&-& 0.142\\
& 2458997.40830	& $-35(8)$ &-& 0.858\\
& 2459010.37335	& $-59(6)$ &-& 0.059\\
& 2459014.60449	& $-53(9)$ &-& 0.451\\
& 2459016.60735	& $-12(9)$ &-& 0.636\\
& 2459017.58234	& $-16(6)$ &-& 0.727\\
& 2460097.77355 & $-17(3)$ &-& 0.763\\
& 2460099.88296	& $-39(3)$ &-& 0.958\\
& 2460103.72427	& $-69(3)$ &-& 0.314\\
& 2460106.73268	& $-19(3)$ &-& 0.593\\
\end{supertabular}
%\end{tabular}
%\end{table}

\begin{figure*}[!t]
\centering

\begin{subfigure}[t]{.40\textwidth}
\centering
\begin{overpic}[width=\linewidth]{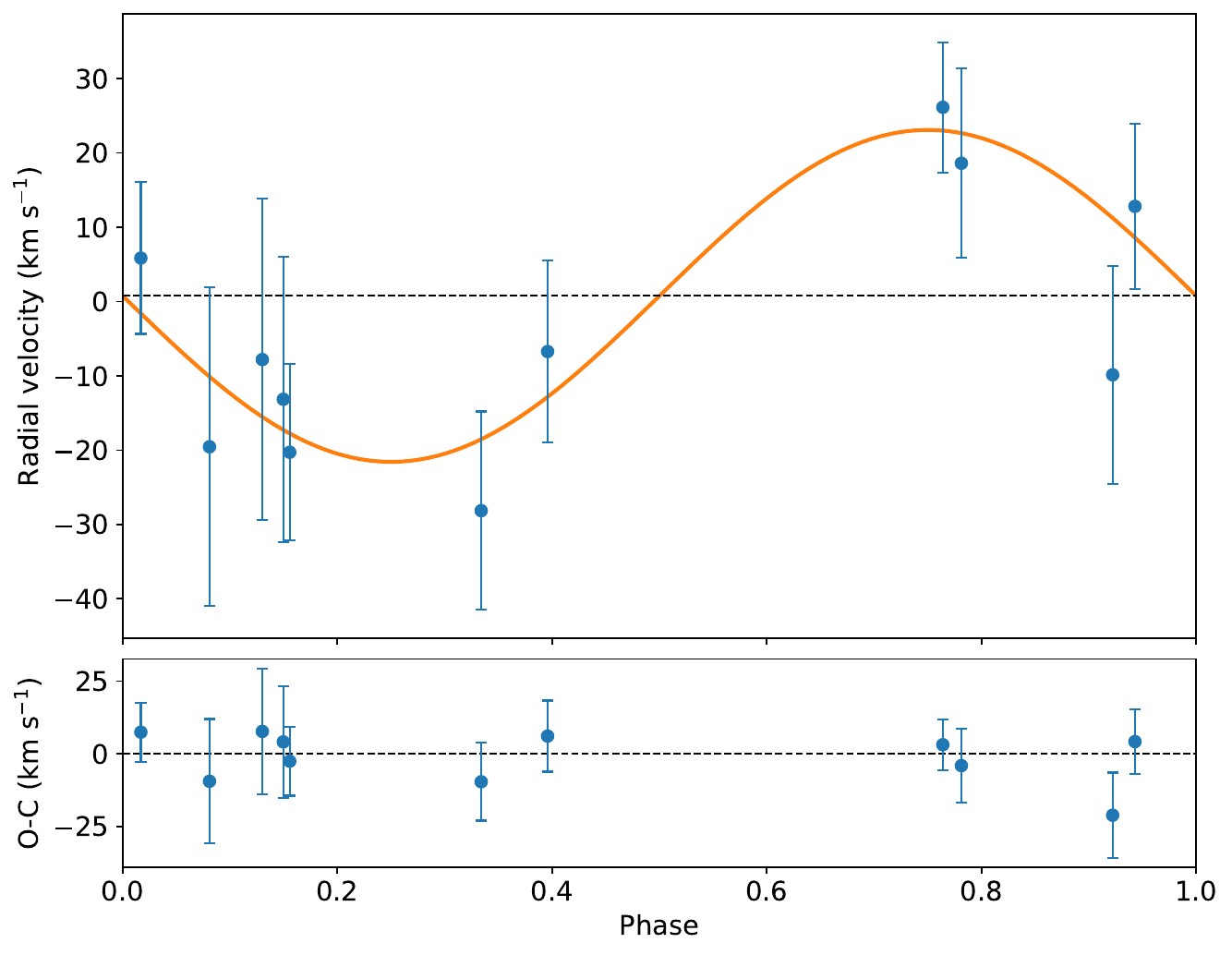}
  \put(15, 65){\small\bfseries HD 101838}
\end{overpic}
\caption{}\label{fig:cpd45}
\end{subfigure}
\begin{subfigure}[t]{.52\textwidth}
\centering
\begin{overpic}[width=\linewidth]{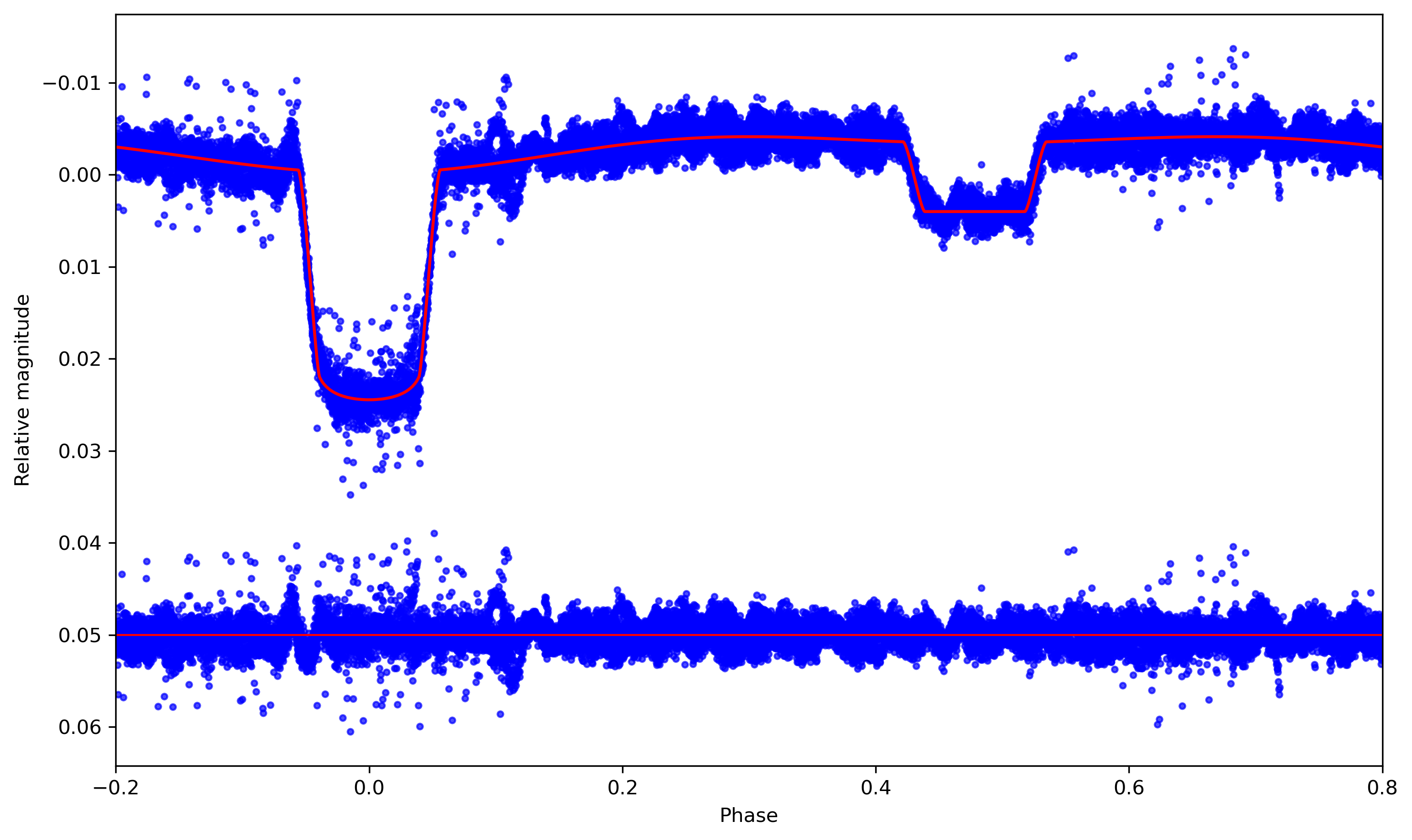}
  \put(40, 30){\small\bfseries HD 101838}
\end{overpic}
\caption{}\label{fig:hd101}
\end{subfigure}

\medskip

\begin{subfigure}[t]{.40\textwidth}
\centering
\begin{overpic}[width=\linewidth]{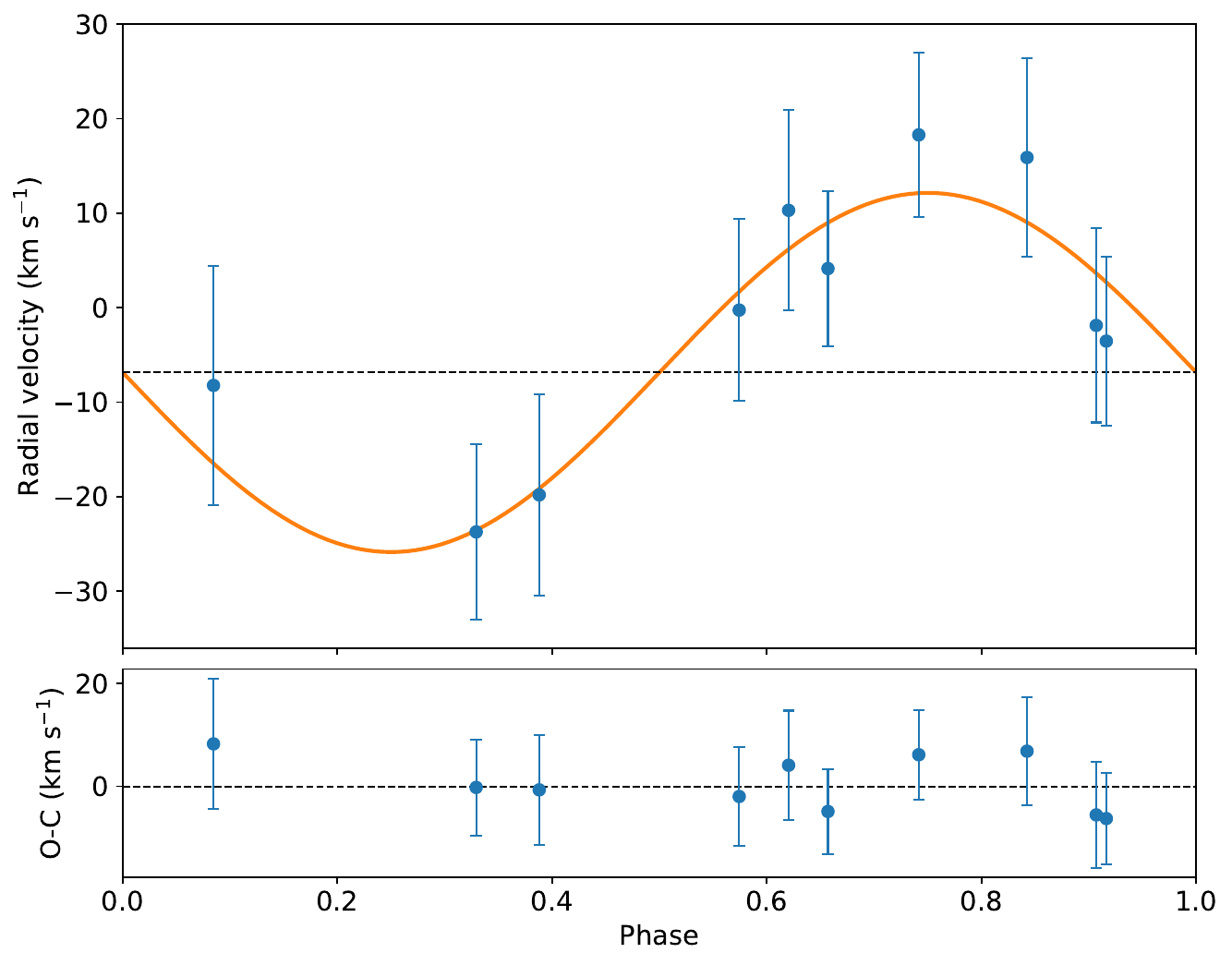}
  \put(15, 70){\small\bfseries HD 108628}
\end{overpic}
\caption{}\label{fig:hd108}
\end{subfigure}
\begin{subfigure}[t]{.52\textwidth}
\centering
\begin{overpic}[width=\linewidth]{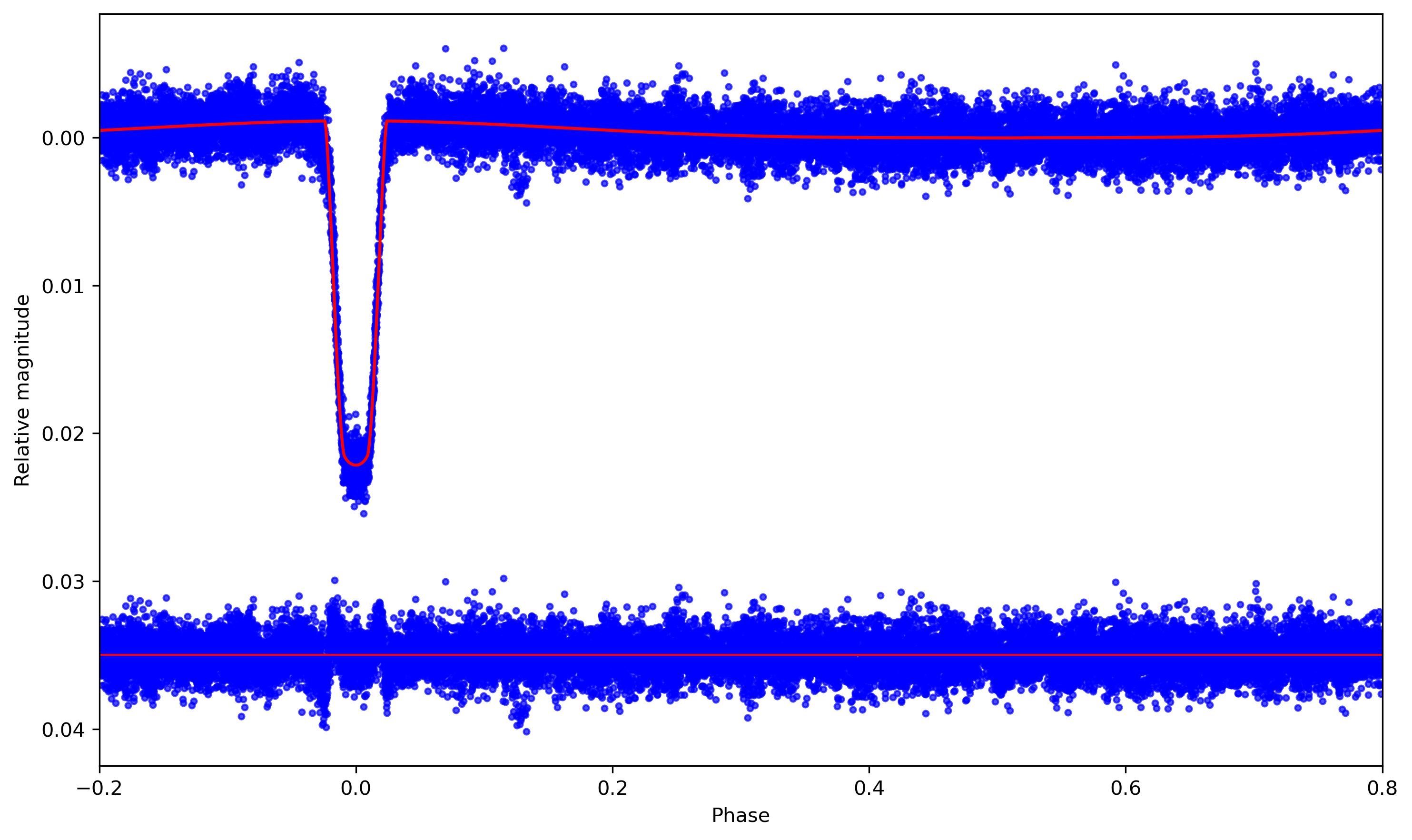}
  \put(40, 30){\small\bfseries HD 108628}
\end{overpic}
\caption{}\label{fig:hd1120}
\end{subfigure}

\medskip

\begin{subfigure}[t]{.40\textwidth}
\centering
\begin{overpic}[width=\linewidth]{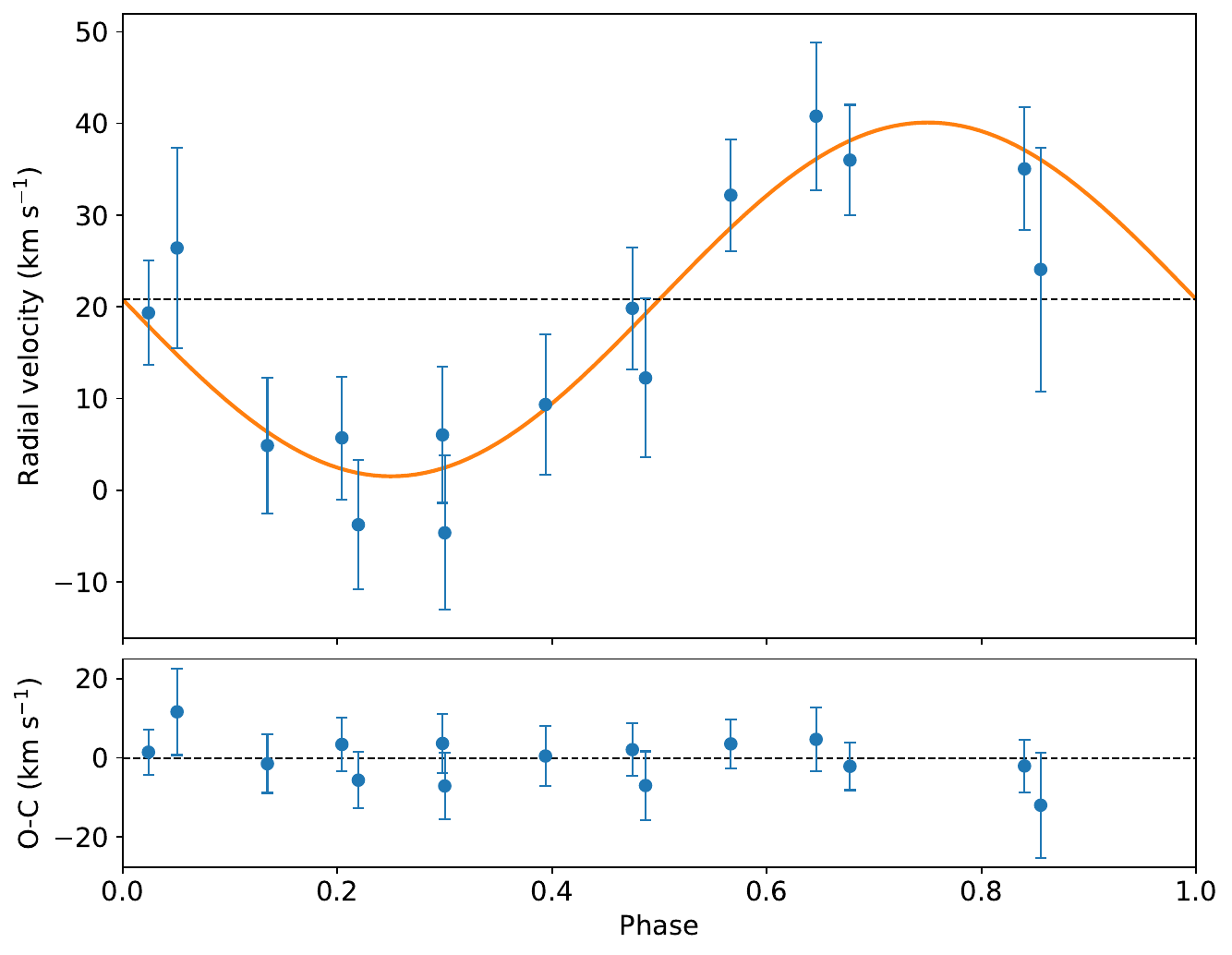}
  \put(15, 70){\small\bfseries HD 254346}
\end{overpic}
\caption{}\label{fig:hd1124}
\end{subfigure}
\begin{subfigure}[t]{.52\textwidth}
\centering
\begin{overpic}[width=\linewidth]{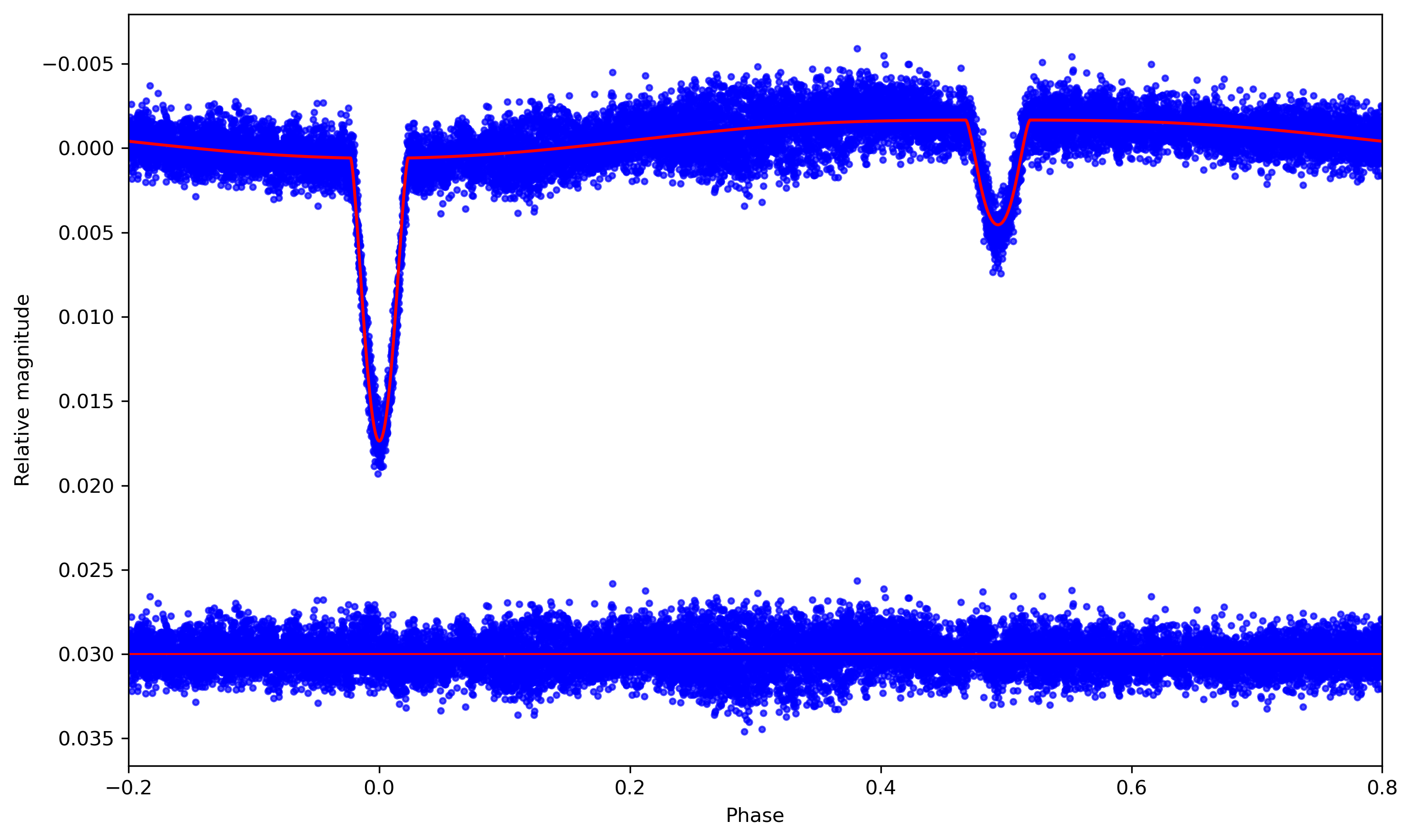}
  \put(40, 30){\small\bfseries HD 254346}
\end{overpic}
\caption{}\label{fig:hd157}
\end{subfigure}

\medskip

\begin{subfigure}[t]{.40\textwidth}
\centering
\begin{overpic}[width=\linewidth]{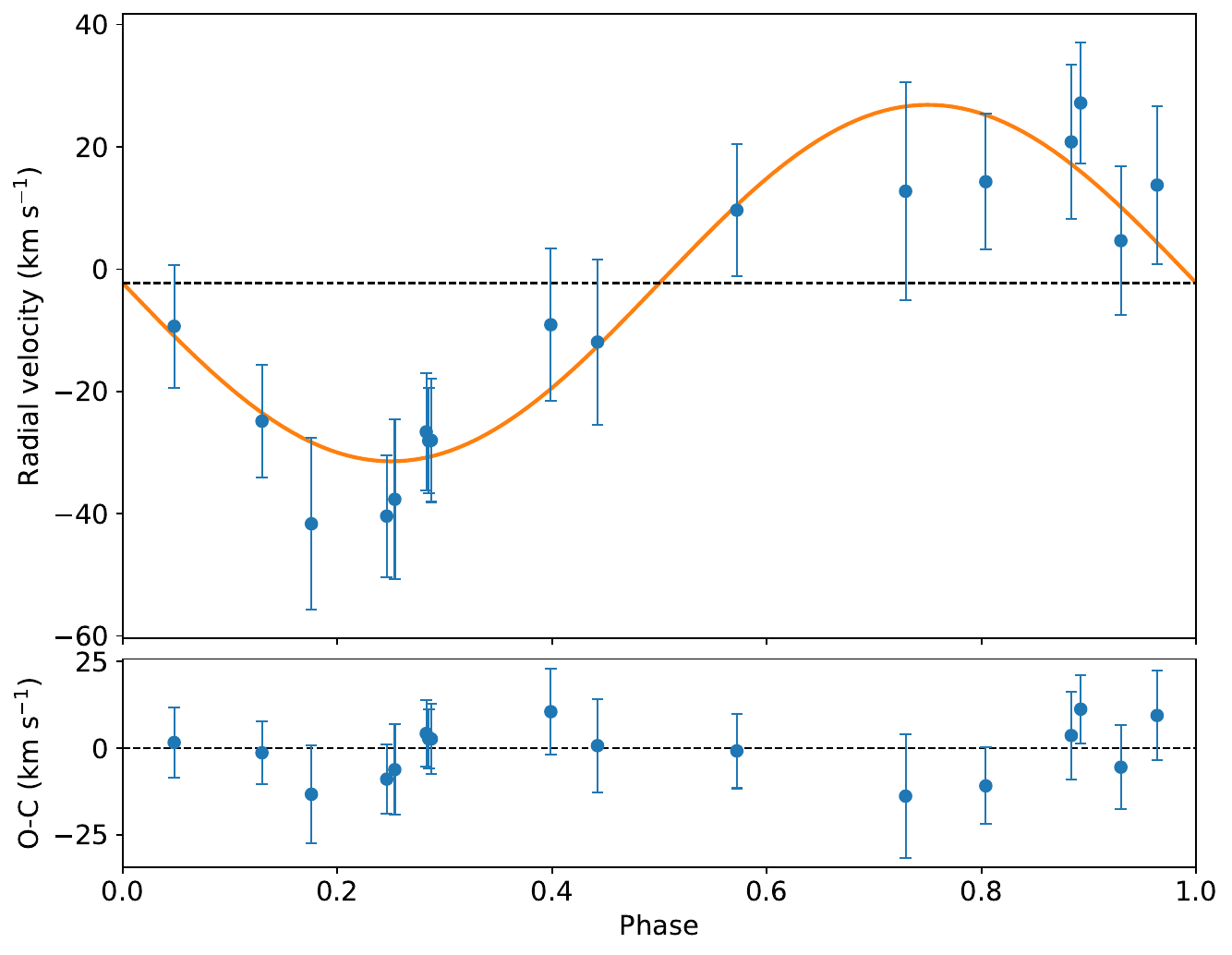}
  \put(15, 70){\small\bfseries HD 339003}
\end{overpic}
\caption{}\label{fig:hd254}
\end{subfigure}
\begin{subfigure}[t]{.52\textwidth}
\centering
\begin{overpic}[width=\linewidth]{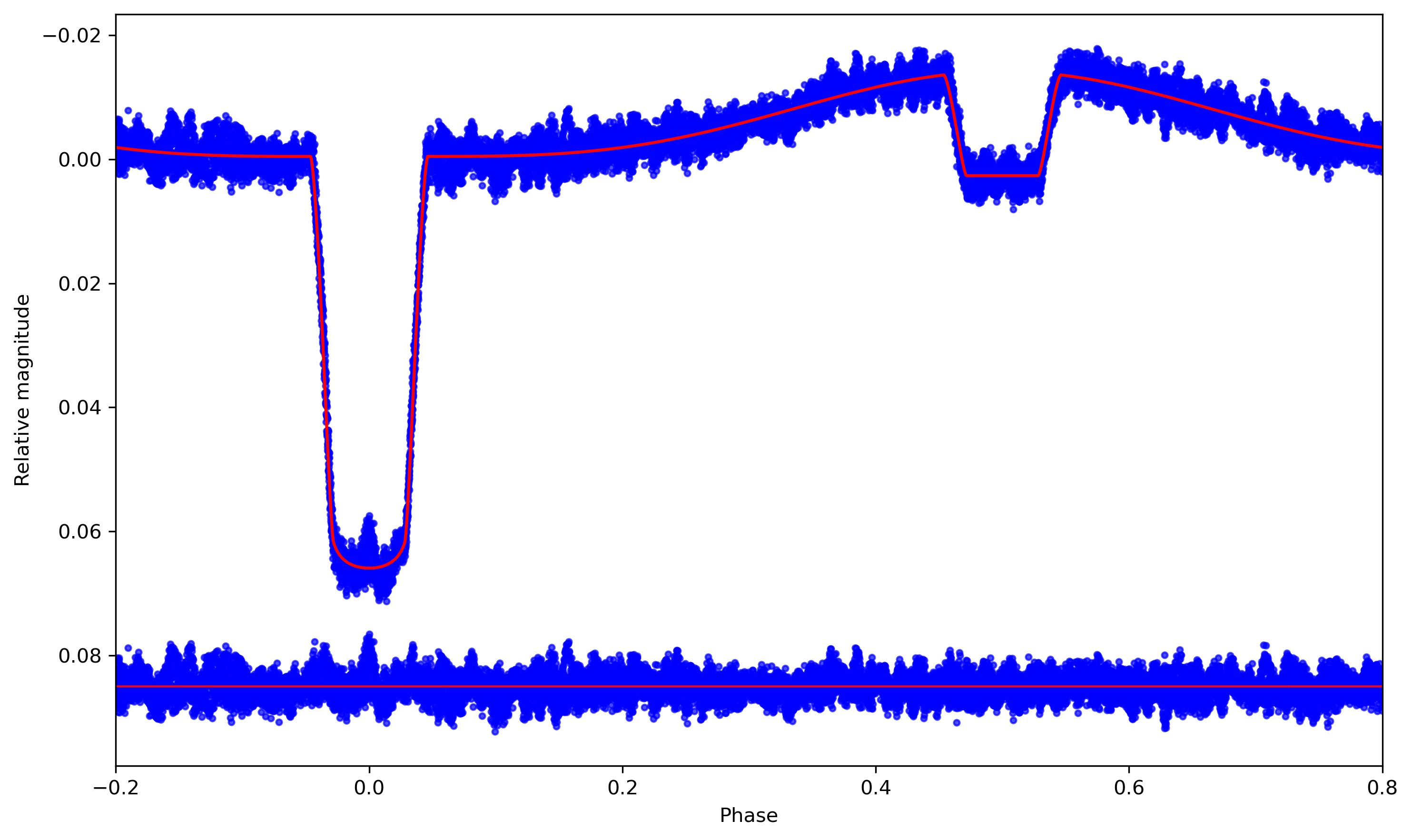}
  \put(40, 30){\small\bfseries HD 339003}
\end{overpic}
\caption{}\label{fig:hd329}
\end{subfigure}
\end{figure*}

\begin{figure*}[!t]
    \ContinuedFloat
    \centering

\begin{subfigure}[t]{.40\textwidth}
\centering
\begin{overpic}[width=\linewidth]{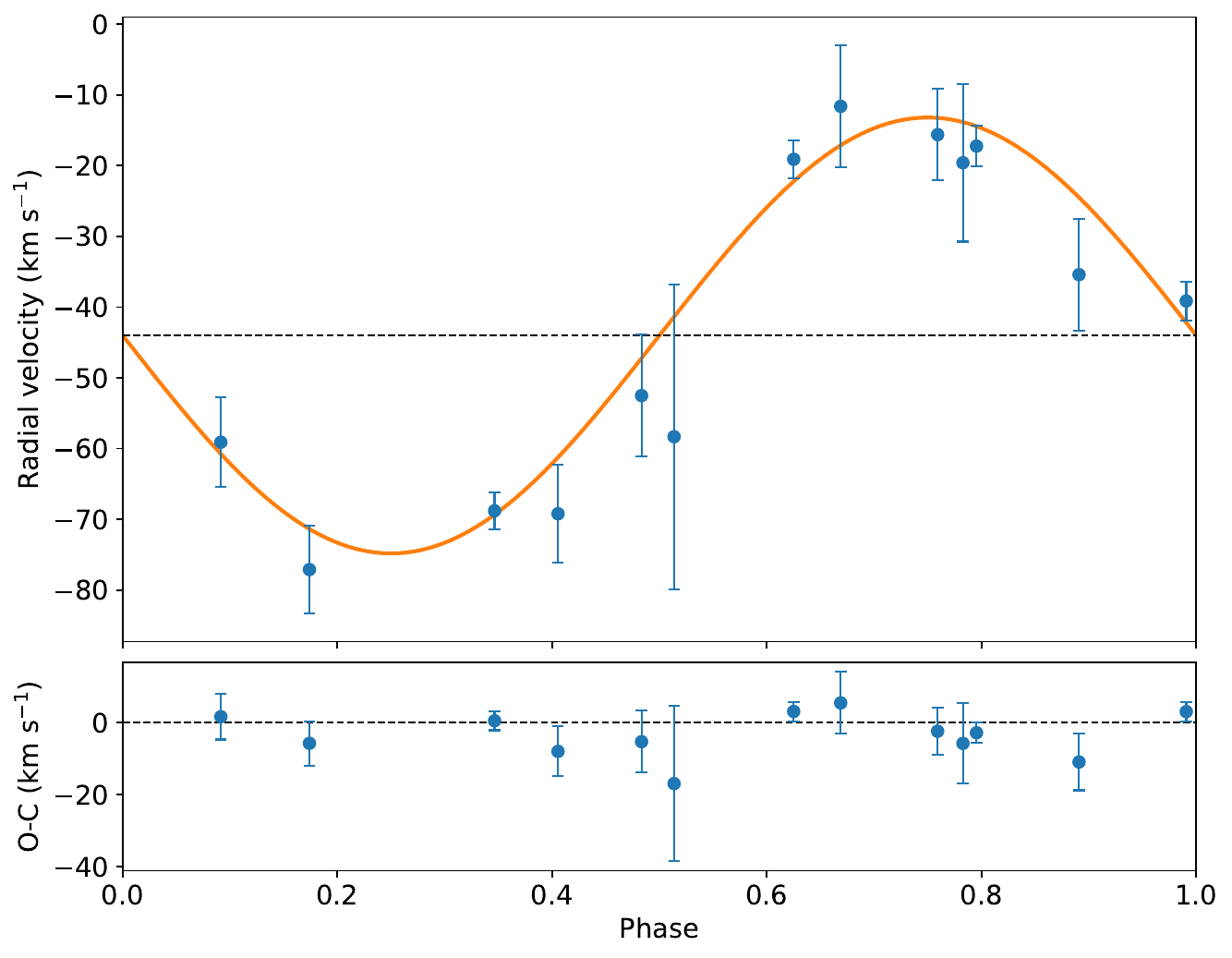}
  \put(15, 70){\small\bfseries V4386 Sgr}
\end{overpic}
\caption{}\label{fig:hd339}
\end{subfigure}
\begin{subfigure}[t]{.52\textwidth}
\centering
\begin{overpic}[width=\linewidth]{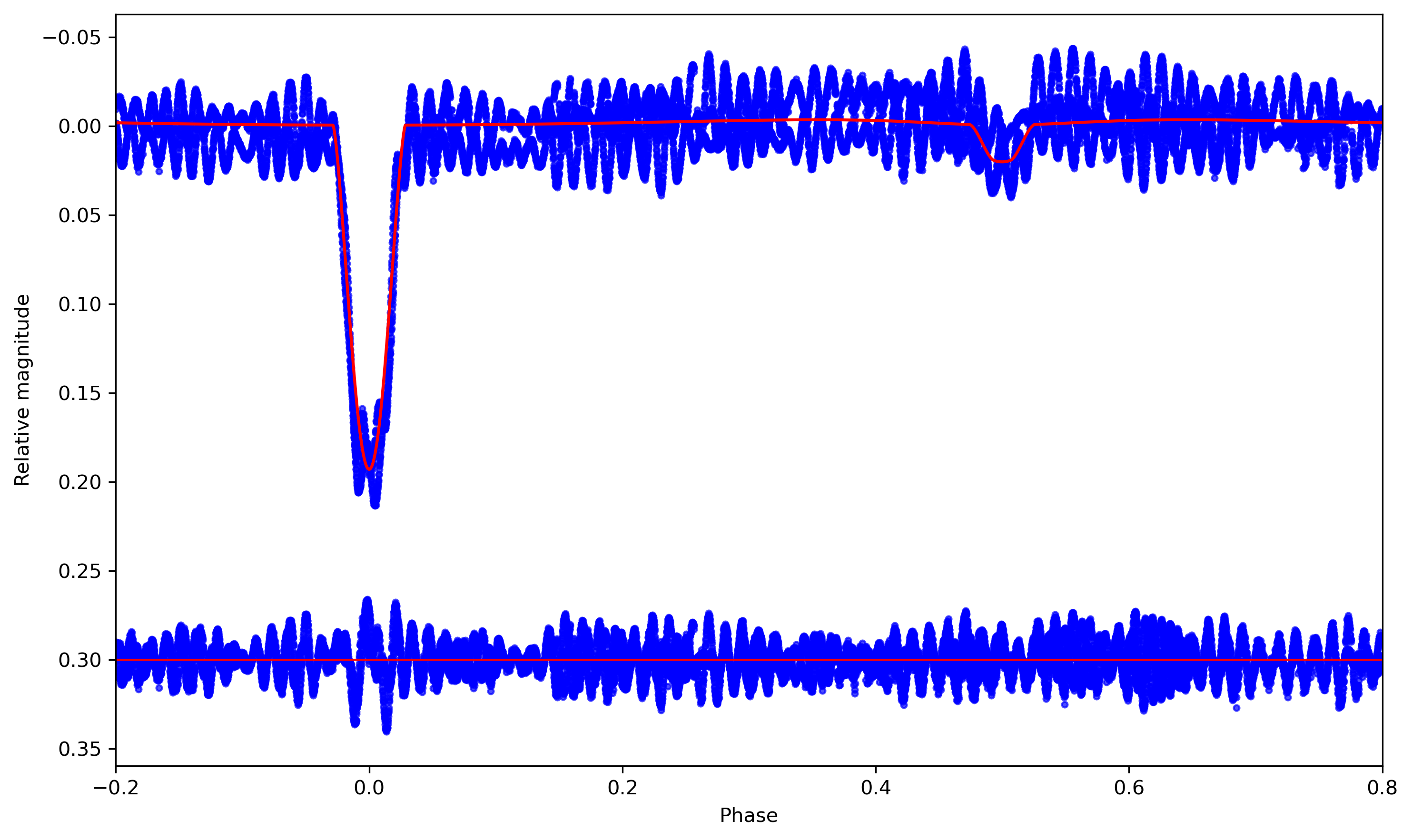}
  \put(40, 30){\small\bfseries V4386 Sgr}
\end{overpic}
\caption{}\label{fig:hd927}
\end{subfigure}

%\begin{minipage}[t]{.4\textwidth}
\caption{Fitted  radial velocity  and light curves of SB1 stars in the sample. }\label{fig:SB1_fits}
%\end{minipage}
\end{figure*}

\begin{figure*}[!b]
\centering
\includegraphics[width=0.85\linewidth]{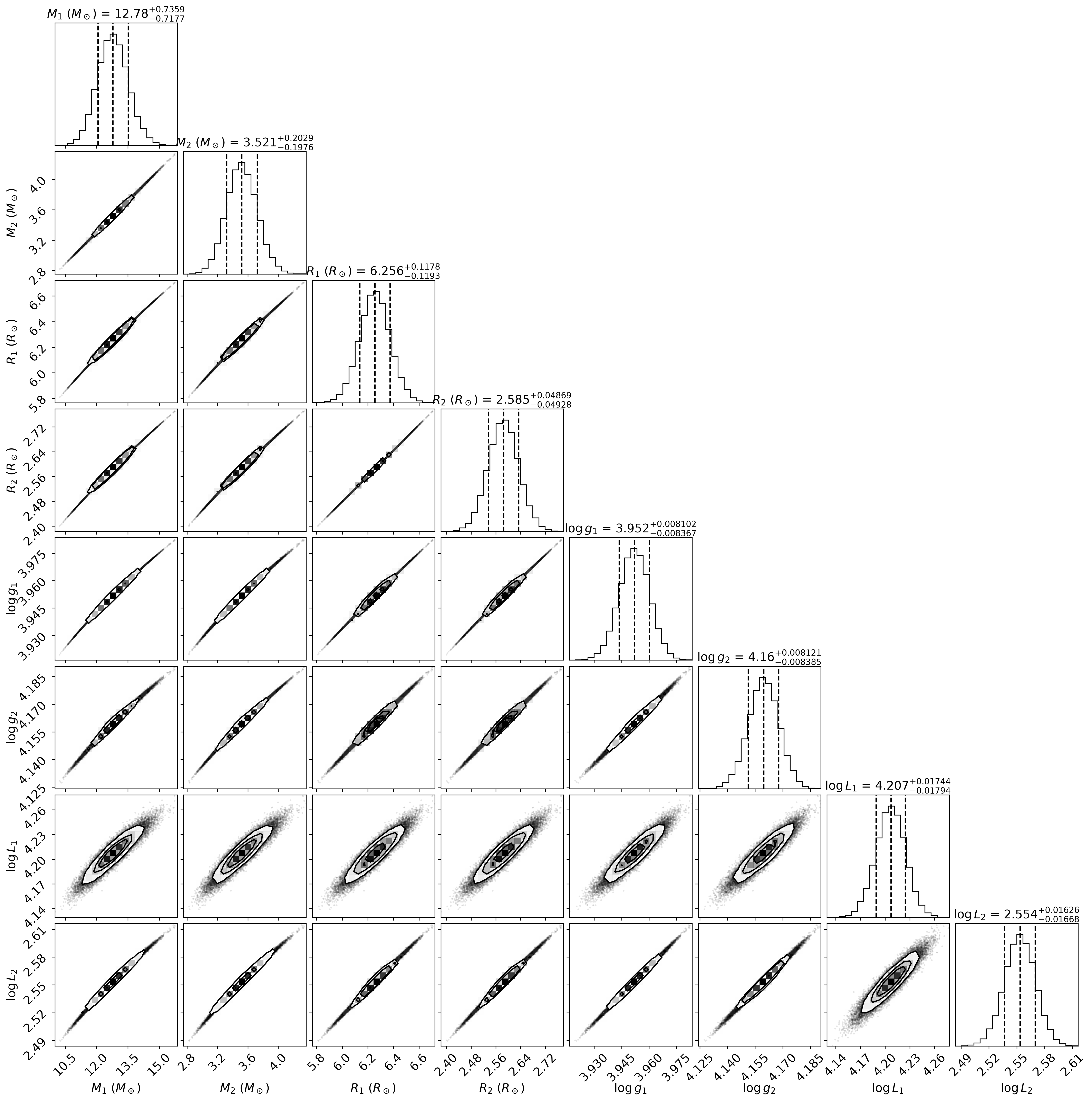}
\caption{An example corner plot showing the derived absolute parameters of V1166 Cen from WD code and their uncertainties. }\label{Appendix:fig_corner_plot_V1166}
\end{figure*}

\section{Fitted atmospheric solutions}\label{Appendix_sec_atm_solns}
Here, we list the line-specific atmospheric parameters derived for the targets. The method for the determination of atmospheric parameters is described in Section \ref{subsec:atmospheric-solution}. * denotes lines that are either too weak or too strong. Such strong or weak lines were excluded from further analysis.

\begin{sidewaystable}
%\begin{table}[ht]
\begin{center}
\caption{The measured $T_{\rm eff}$ and $\log g$ from individual lines . }\label{tab:Teffandlogg-from-lines}
%\begin{adjustbox}{angle=90}
\begin{tabular}[t]{llccccccc}
\hline\hline
Star	&	${T_{\rm eff}\_\rm H\delta \, K}$	&	$\log g\_\rm H\delta$	&	${T_{\rm eff}\_\rm H\beta \, K}$ &	$\log g\_\rm H\beta$	&	${T_{\rm eff}\_\rm H\alpha \, K}$ &	$\log g\_\rm H\alpha$ &	${T_{\rm eff}\_\rm He\,{\sc i} \, K}$	&	$\log g\_\rm He\,{\sc i}\, K$	\\
CD$-38$ 4128	&	21500(1000)	&	3.7(1)	&	21000(1000)	&	3.7(2)	&		&		&	20000(2000)*	&	3.7(2)	\\
CPD$-45$ 3109	&	21500(500)	&	3.7(1)	&	22000(1000)	&	3.5(2)	&		&		&	22000(2000)	&	3.5(2)	\\
HD101838	&	27000(2000)	&	3.7(2)	&	27000(2000)	&	3.5(2)	&		&		&		&		\\
HD 108628	&	25500(1500)	&	4.0(3)	&	24000(3000)	&	3.9(1)	&		&		&	24000(3000)	&	3.9(1)	\\
HD112026	&		&		&		&		&	28000(1500)	&	3.5(2)	&		&		\\
HD 112485$_{1}$	&		&		&	25500(1500)	&	3.9(2)	&		&		\\
HD 157400	&	23000(1000)	&	4.0(1)	&	24000(2000)	&	3.85(20)	&		&		&	20000(3000)*	&	4.0(3)	\\
HD 254346	&	24000(1500)	&	3.9(2)	&	26000(2000)	&	3.75(20)	&		&		&		&		\\
HD 329379$_{1}$	&		&	&	25000(2000)	&	3.75(20)	&		&		&	&	\\
HD 329379$_{2}$	&	&	&	15000(2000)	&	3.7(2)	&		&		&	&\\
HD 339003	&	28000(1500)	&	3.85(2)	&	29000(2500)	&	3.75(20)	&	29000(-)	&	3.85(2)	&		&		\\
v1166 Cen$_{1}$	&		&		&	26000(2000)	&	3.9(2)	&		&		&	&	\\
V4386 Sgr	&	23500(1000)	&	3.6(1)	&	24000(1500)	&	3.5(2)	&		&		&	24000(1000)	&	3.8(3)	\\
\end{tabular}
%\end{adjustbox}
\end{center}
%\end{table}
\end{sidewaystable}

\section{Targets with Data limitations}\label{appendix:sec_data_limitations}
Here, we discuss four systems with varying degree of data limitations. HD 112026 and HD 157400 are highly eccentric SB2 whose disentangling was marred by a limited number of spectra. As a result, it was not possible to reliably derive plausible $K_{2}$ values of their secondary components and subsequently their mass ratio and semi-major axis. HD 112026 and HD 92741 had artefacts in the Balmer lines such that it was not possible to derive the effective temperature of their primary component from those lines, coupled with their limited number of spectra. CPD$-$45 3109 on the other hand, does not have 2-min TESS light curves. For these systems, we report only their orbital parameters derived from their spectra and the inclination derived from their preliminary  light curve model on a case-by-case basis using appropriate software such as the WD code \citep{WilsonandDevinney1971}, JKTEBOP \citep{Southworthetal2004} and \textsc{STAR SHADOW} \citep{Ijspeertetal2024}.

\textsc{STAR SHADOW} is a Fourier based python software that automatically derives the physical parameters such as inclination, sum of radii, fractional radii, surface brightness ratio etc. using eclipse timing and binary harmonic models. It builds binary harmonic model by modelling eclipses and sinusoids using the harmonic frequencies it extracts from the light curve. \textsc{STAR SHADOW}  uses the orbital period as input and automatically extracts the harmonic frequencies through successive prewhitening. See \citet{Ijspeertetal2024} for details on the use of \textsc{STAR SHADOW}. 

\begin{figure*}[!b]
\centering
\begin{subfigure}[t]{.42\textwidth}
\centering
\begin{overpic}[width=\linewidth]{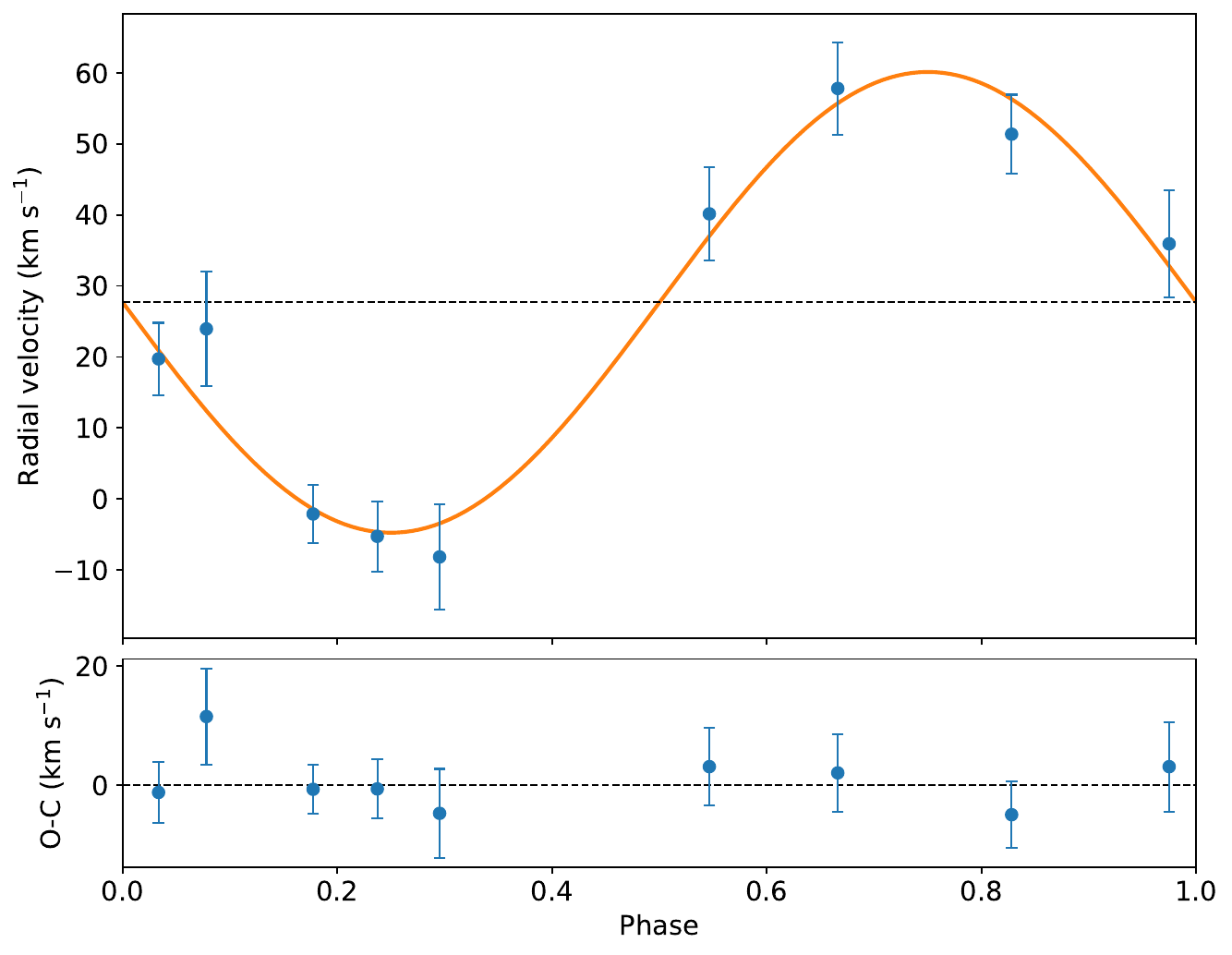}
  \put(12, 70){\small\bfseries CPD$-$45 3109}
\end{overpic}
\caption{}\label{fig:cpd45}
\end{subfigure}
\begin{subfigure}[t]{.405\textwidth}
\centering
\begin{overpic}[width=\linewidth]{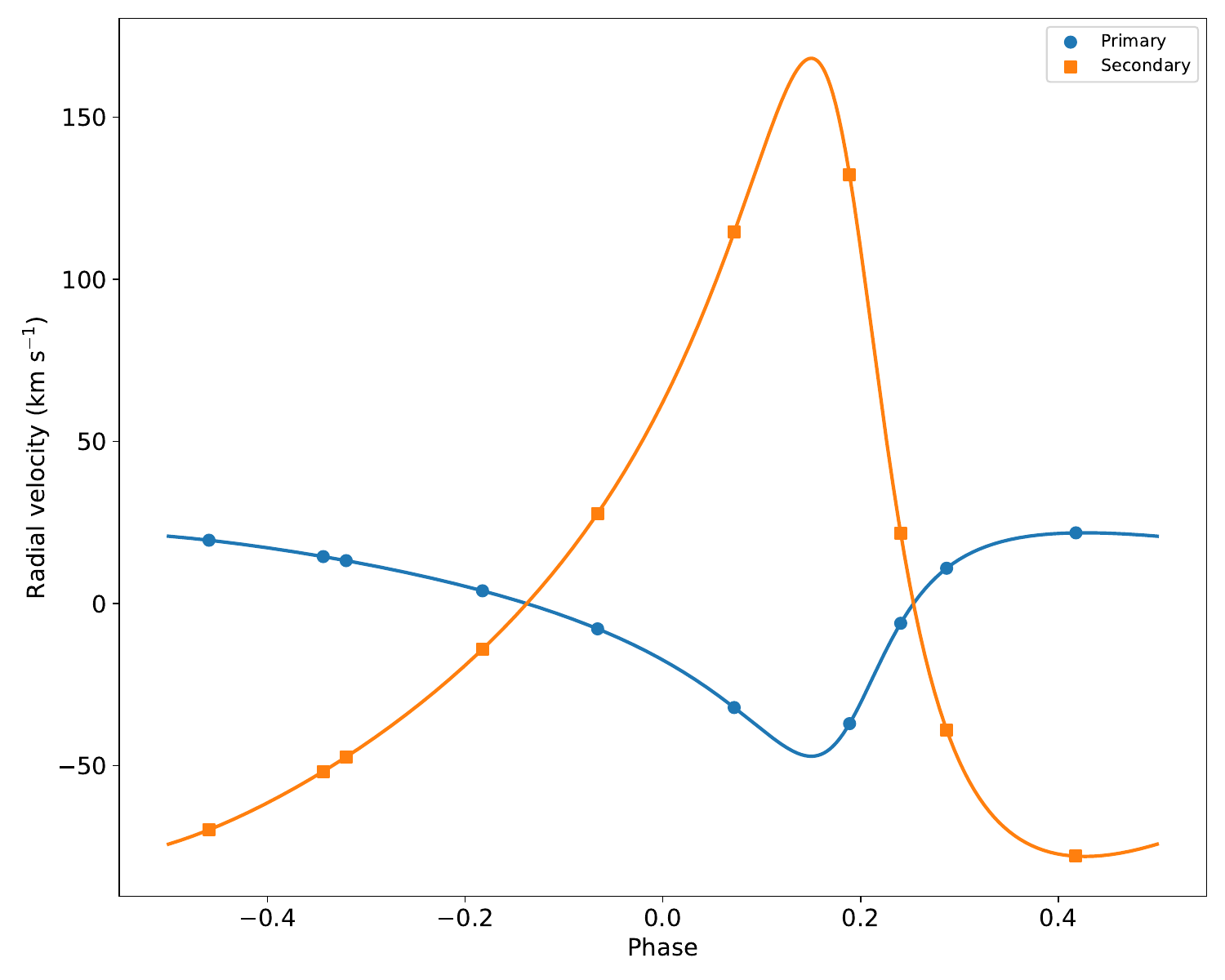}
  \put(20, 52){\small\bfseries HD 112026}
\end{overpic}
\caption{}\label{fig:cpd45}
\end{subfigure}

\medskip

\begin{subfigure}[t]{.45\textwidth}
\centering
\begin{overpic}[width=\linewidth]{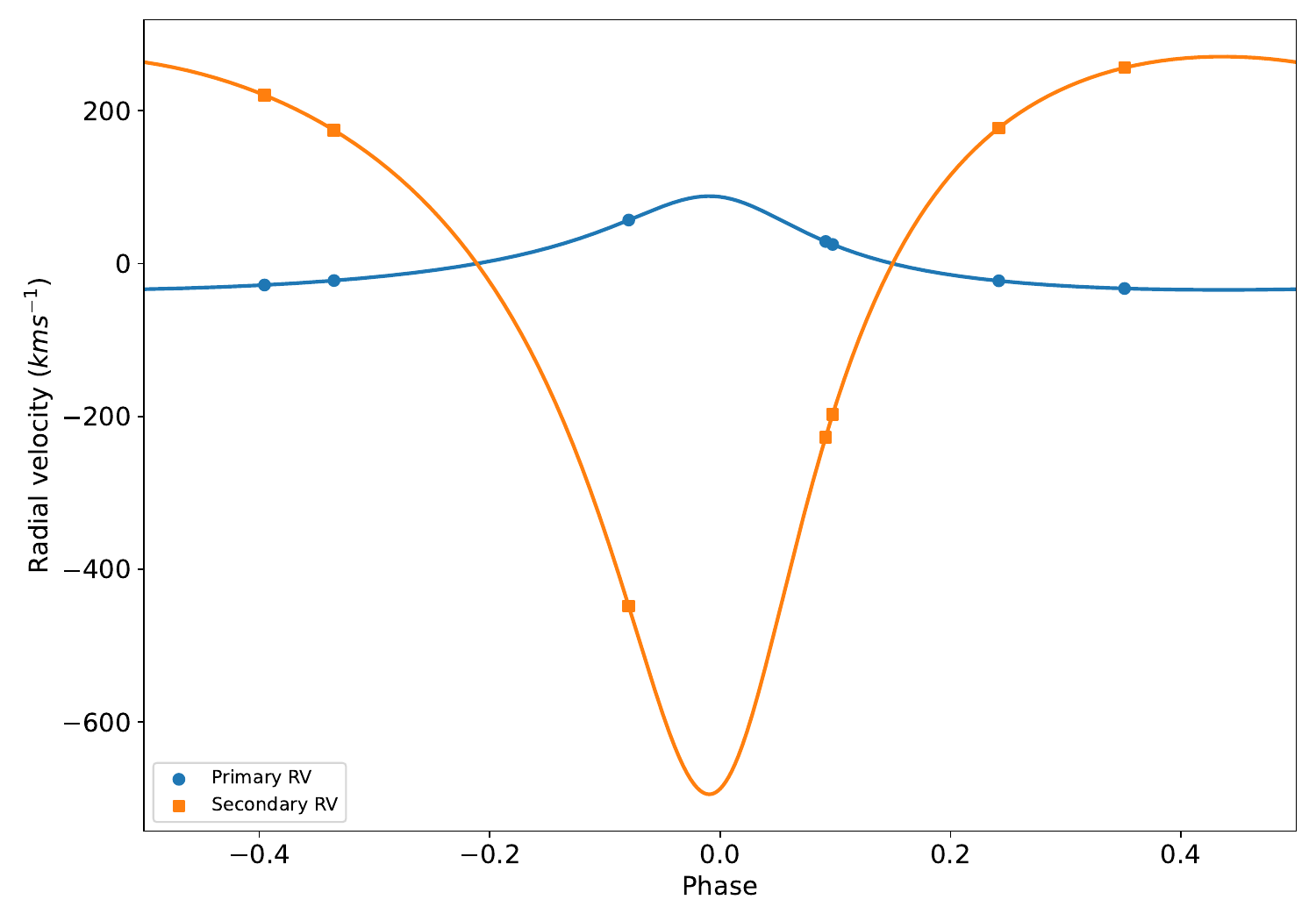}
  \put(50, 60){\small\bfseries HD 157400}
\end{overpic}
\caption{}\label{fig:hd101}
\end{subfigure}
\begin{subfigure}[t]{.40\textwidth}
\centering
\begin{overpic}[width=\linewidth]{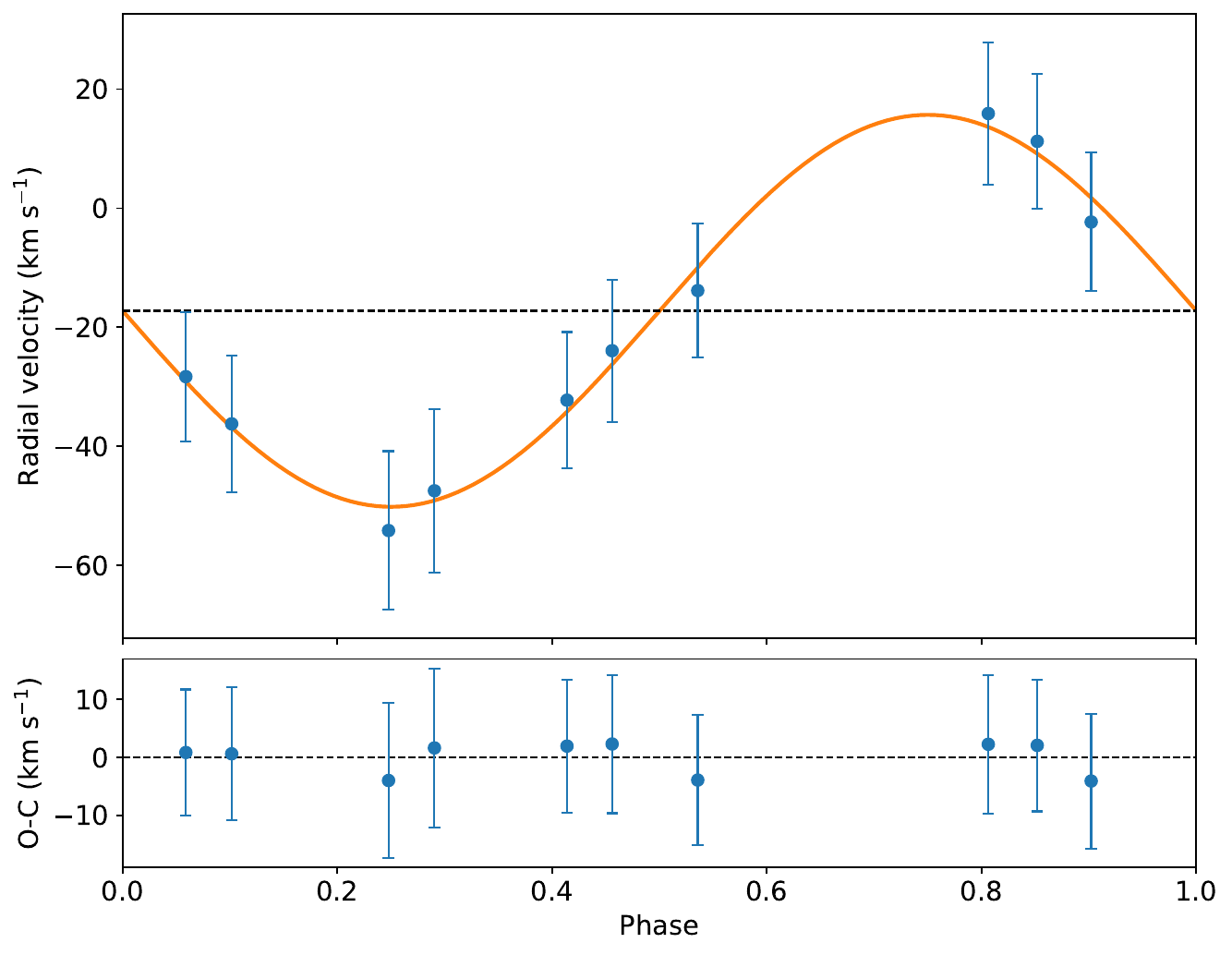}
  \put(55, 40){\small\bfseries HD 92741}
\end{overpic}
\caption{}\label{fig:hd1120}
\end{subfigure}
\caption{Fitted  radial velocity  of the four systems with data limitations. }\label{fig:RVfits_for_systems_data_limitation}
\end{figure*}

The inclination of HD 112026 was estimated using the WD code and that of HD 157400 was estimated using JKTEBOP in line with Sections \ref{sec:SB2} and \ref{sec:SB1}, respectively. Those of CPD$-45$ 3109 and HD 92741 were automatically derived using \textsc{STAR SHADOW}. The QLP light curve \citep{Huangetal2020a, Huangetal2020b} was used for the estimation of inclination in the case of CPD$-45$ 3109. Orbital solutions of HD 112026 and HD 157400 were derived via disentangling and those of CPD$-45$ 3109 and HD 92741 were derived using \textsc{RVFIT} in line with Section \ref{sec:SB1}. Figure \ref{fig:RVfits_for_systems_data_limitation} shows the fitted RV curves of CPD$-45$ 3109 and HD 92741  and Table \ref{Appendix:tab_orbital_elements_data_limitaions} shows the fitted orbital parameters of the systems with data limitations. 

Given the parameters of HD~157400, the semi-major axis of approximately 163~$R_{\odot}$ and M$_{1}$ of 180~$M_{\odot}$ are expected to be obtained. However, considering the spectral type of the star, these values of $a$ and $M_{1}$ are grossly overestimated and unphysical. A good explanation is that the $K_{2}$ value is grossly overestimated, resulting in the enormous orbital scale and mass. On the other hand, the $K_{2}$ value for HD~112026 appears to be underestimated as it suggests  semi-major axis of 118~$R_{\odot}$ and M$_{1}$ of 9~$M_{\odot}$, which do not corroborate the spectral class and the observed $T_{\rm eff}$. 

For placing HD 112026 and HD 157400 in the HR diagram (Figure \ref{fig:HR_diagram}) and synchronisation plot (Figure \ref{fig:P_Prot_plot}), the rough estimates of their primary  masses, radii, ages, luminosities and rotational periods were derived in line with Sections \ref{subsec:absolute-parameter} and \ref{subsec:circularisation-and-synchronisation} using parameters from their non-disentangled spectra. The mass, radius, age, $\log L$, and $P_{r}$ of $9.8 (4)\,\rm M_{\odot}$, $5.4 (5)\,\rm R_{\odot}$, $13.8 (3.4)\,\rm Myr$, $3.91 (5)\,\rm L_{\odot}$ and $1.4 (2)\,\rm d$, respectively, are obtained for HD 157400. For HD 112026, $15.4 (9)\,\rm M_{\odot}$, $9 (1)\,\rm R_{\odot}$, $8.6 (1.1)\,\rm Myr$, $4.61 (8)\,\rm L_{\odot}$ and $7 (1)\,\rm d$ are derived respectively for mass, radius, age, $\log L$, and $P_{r}$. However, as observed in Figure \ref{fig:fig_logg_logg}, the derived parameters for HD 112026 could not adequately reproduce the observed $\log g$. This could be partly due to artefacts in the Balmer lines that impaired the determination of atmospheric parameters.

\begin{table}[ht]
\begin{center}
\caption{Fitted parameters four remaining targets in our sample with varying limitations. "f" denotes fixed parameters.}\label{Appendix:tab_orbital_elements_data_limitaions}
\begin{tabular}[t]{ccc}
\hline
Star   & CPD$-$45 3109   & HD 92741\\
\hline\hline
Adjusted Quantities\\
$P$ (d)	&$2.7139^{f}$   & $5.373^{f}$ \\
$T_p$ (HJD$-$2457000)	&$2919.50^{f}$ & $1574.45^{f}$ \\
$e$			 &0.0(1)   & $0^{f}$ \\
$\omega$ ($^\circ$)	 &67(6)   &90  \\
$i$ ($^\circ$)     &61(4)  & 74(2)  \\
$\gamma$ ($\rm km\,s^{-1}$)	& 28(2)  &$-17(4)$ \\
$K_1$ ($\rm km\,s^{-1}$)		&32(3)   &  33(6) \\
$K_2$ ($\rm km\,s^{-1}$)	& - & -\\
\hline
Derived Quantities\\
$a_1\sin i$ ($10^6$ km)	 &1.2(1)  & 2.4(4) \\
$f(m_1,m_2)$ ($M_\odot$)	 &0.009(5)  & 0.02(1)  \\
\hline
Other Quantities\\
$\chi^2$		&3.84   & 0.49 \\
%$N_{obs}$ (primary)	 &9  &10  &7 &  10\\
$rms_1$ ($\rm km\,s^{-1}$) &4.78   & 2.64 \\
\hline\hline
Star     & HD 112026  & HD 157400  \\
\hline\hline
Adjusted Quantities\\
$P$ (d)	&$43.205^{f}$  &$16.9045^{f}$  \\
$T_p$ (HJD$-$2457000)	 &2356.57$^{f}$  &$3101.20^{f}$  \\
$e$			   & 0.49(4)&0.445(5)  \\
$\omega$ ($^\circ$)	   &222(9)  &9.95(5)  \\
$i$ ($^\circ$)      & 86.21(4)  &89.135(58)   \\
$\gamma$ ($\rm km\,s^{-1}$)	 & - & -  \\
$K_1$ ($\rm km\,s^{-1}$)		  & 34.5(9) &61(11)  \\
$K_2$ ($\rm km\,s^{-1}$)	 &123(53) &483(57)  \\
\hline
Derived Quantities\\
$a_1\sin i$ ($10^6$ km)	   &-  & -  \\
$f(m_1,m_2)$ ($M_\odot$)	   &-  &-  \\
\hline
Other Quantities\\
$\chi^2$		 &-  &-  \\
%$N_{obs}$ (primary)	 &9  &10  &7 &  10\\
$rms_1$ ($\rm km\,s^{-1}$)   & -&  - \\
\hline\hline
%\\
%Table \ref{tab:orbital-element} continues\\
\end{tabular}
\end{center}
\end{table}

\end{appendix}

\end{document}